\documentclass[prx,aps,twocolumn,superscriptaddress,showpacs,notitlepage,longbibliography,10pt]{revtex4-1}

\usepackage{amsmath,amssymb,amsfonts,amstext}
\usepackage{graphicx,breakurl,color,setspace}
\usepackage{mathtools}
\usepackage{dsfont}
\usepackage[T1]{fontenc}
\usepackage[latin9]{inputenc}
\usepackage[normalem]{ulem}
\usepackage[unicode=true,pdfusetitle,pdfborder={0 0 1}]{hyperref}
\usepackage{upgreek}
\usepackage{bm}
\usepackage{mathrsfs}

\hypersetup{colorlinks,linkcolor=blue,citecolor=blue,filecolor=blue,urlcolor=blue}

\newcommand{\braket}[2]{\langle #1 | #2 \rangle} 
\newcommand{\bra}[1]{\langle #1 |} 
\newcommand{\ket}[1]{| #1 \rangle} 
\newcommand{\bracket}[3]{\langle #1 | #2 | #3 \rangle} 
\newcommand{\ave}[1]{\left< #1 \right>} 
\newcommand{\ketbra}[1]{|#1\rangle\langle#1|}

\newcommand{\etal}{\textit{et al.}}
\newcommand{\vect}[1]{\boldsymbol{#1}}
\newcommand{\Id}{\mathds{1}}
\DeclareMathOperator{\poly}{poly}
\DeclareMathOperator{\Var}{Var}

\newcommand{\vgamma}{\vec{\bm\gamma}}
\newcommand{\vbeta}{\vec{\bm\beta}}
\newcommand{\vu}{\vec{\bm u}}
\newcommand{\vv}{\vec{\bm v}}

\newcommand{\z}{\vect{z}}
\newcommand{\cP}{\mathcal{P}}

\begin{document}

\title{Quantum Approximate Optimization Algorithm: Performance, Mechanism, and Implementation on Near-Term Devices}

\author{Leo Zhou}
\email{leozhou@g.harvard.edu}
\affiliation{Department of Physics, Harvard University, Cambridge, MA 02138, USA }

\author{Sheng-Tao Wang}
\email{shengtaowst@gmail.com}
\thanks{\\L.Z. and S.-T.W. contributed equally to this work.}
\affiliation{Department of Physics, Harvard University, Cambridge, MA 02138, USA }

\author{Soonwon Choi}
\affiliation{Department of Physics, Harvard University, Cambridge, MA 02138, USA }
\affiliation{Department of Physics, University of California Berkeley, Berkeley, CA 94720, USA }

\author{Hannes Pichler}
\affiliation{ITAMP, Harvard-Smithsonian Center for Astrophysics, Cambridge, MA 02138, USA }
\affiliation{Department of Physics, Harvard University, Cambridge, MA 02138, USA }

\author{Mikhail D. Lukin}
\affiliation{Department of Physics, Harvard University, Cambridge, MA 02138, USA }

\date{\today}

\begin{abstract}
The Quantum Approximate Optimization Algorithm (QAOA) is a hybrid quantum-classical variational algorithm designed to tackle combinatorial optimization problems.
Despite its promise for near-term quantum applications, not much is currently understood about QAOA's performance beyond its lowest-depth variant.
An essential but missing ingredient for understanding and deploying QAOA is a constructive approach to carry out the outer-loop classical optimization.
We provide an in-depth study of the performance of QAOA on MaxCut problems by developing an efficient parameter-optimization procedure and revealing its ability to exploit non-adiabatic operations.
%
%
Building on observed patterns in optimal parameters, we propose heuristic strategies for initializing optimizations to find quasi-optimal $p$-level QAOA parameters in $O(\poly(p))$ time, whereas the standard strategy of random initialization requires $2^{O(p)}$ optimization runs to achieve similar performance. 
We then benchmark QAOA and compare it with quantum annealing, especially on difficult instances where adiabatic quantum annealing fails due to small spectral gaps.
The comparison reveals that QAOA can learn via optimization to utilize non-adiabatic mechanisms to circumvent the challenges associated with vanishing spectral gaps. 
Finally, we provide a realistic resource analysis on the experimental implementation of QAOA.
When quantum fluctuations in measurements are accounted for,
we illustrate that optimization will be important only for
problem sizes beyond numerical simulations, but accessible on near-term devices.
We propose a feasible implementation of large MaxCut problems with a few hundred vertices in a system of 2D neutral atoms, reaching the regime to challenge the best classical algorithms.
\end{abstract}

\maketitle

\section{Introduction}

As quantum computing technology develops, 
there is a growing interest in finding useful applications of near-term quantum machines~\cite{Preskill2018Quantum}.
In the near future, however, the number of reliable quantum operations will be limited by noise and decoherence. 
As such, hybrid quantum-classical algorithms~\cite{QAOA, Peruzzo2014, Moll2018Quantum} have been proposed to make the best of available quantum resources and integrate them with classical routines.
The Quantum Approximate Optimization Algorithm (QAOA) \cite{QAOA} and the Variational Quantum Eigensolver  \cite{Peruzzo2014} are such algorithms put forward to address classical combinatorial optimization and quantum chemistry problems, respectively.
Proof-of-principle experiments running these algorithms have already been demonstrated in the lab~\cite{Kandala2017Hardware, Otterbach2017Unsupervised, Kokail2018Self, Qiang2018Photonics}.

In these hybrid algorithms, a quantum processor prepares a quantum state according to a set of variational parameters.
Using measurement outputs, the parameters are then optimized by a classical computer and fed back to the quantum machine in a closed loop. In QAOA, the state is prepared by a $p$-level  circuit specified by $2p$ variational parameters.
Even at the lowest circuit depth~($p=1$), QAOA has non-trivial provable performance guarantees~\cite{QAOA, QAOAE3lin} and is not efficiently simulatable by classical computers~\cite{QAOAsupremacy}.
It is thus an appealing algorithm to explore quantum speedups on near-term quantum machines.

However, very little is known about QAOA beyond the lowest depth.
While QAOA is known to monotonically improve with depth and succeed in the $p\to\infty$ limit~\cite{QAOA}, its performance when $1<p<\infty$ is largely unexplored.
In fact, it has been argued that one needs to go beyond low-depth QAOA in order to compete with the best classical algorithm for some problems on bounded-degree graphs~\cite{Hastings2019Classical, BKKT2019}.
It thus remains a critical problem to assess QAOA at intermediate depths where one may hope for a quantum computational advantage.
One major hurdle lies in the difficulty to efficiently optimize in the non-convex, high-dimensional parameter landscape.
Without constructive approaches to perform the parameter optimization, any potential advantages of the hybrid algorithms could be lost~\cite{McCleanBarrenLandscape}.

In this work, we contribute, in three major aspects, to the understanding and applicability of QAOA on near-term devices, with a focus on MaxCut problems.
First, we develop heuristic strategies to efficiently optimize QAOA variational parameters.
These strategies are found, via extensive benchmarking, to be quasi-optimal in the sense that they usually produce known global optima. The standard approach with random initialization generically require $2^{O(p)}$ optimization runs to surpass our heuristics.
Secondly, we benchmark the performance of QAOA and compare it with quantum annealing.
On difficult graph instances where the minimum spectral gap is very small, the time required for quantum annealing to remain adiabatic is very long as it scales inversely with the square of the gap.
For these instances, QAOA is found to outperform adiabatic quantum annealing by multiple orders of magnitude in computation time.
Lastly, we provide a detailed resource analysis on the experimental implementation of QAOA with near-term quantum devices. Taking into account of quantum fluctuations in projective measurements, we argue that optimization will play a role only for much larger problem sizes than numerically accessible ones. We also propose a 2D physical implementation of QAOA on MaxCut with a few hundred Rydberg-interacting atoms, which can be put to the test against the best classical algorithm for potential quantum advantages.  
		
Our main results can be summarized as follows.
By performing extensive searches in the entire parameter space, we discover persistent patterns in the optimal parameters.
%
%
Based on the observed patterns, we develop strategies
for selecting initial parameters in optimization,
which allow us to efficiently optimize QAOA at a cost scaling polynomially in $p$.
This is in stark contrast to the $2^{O(p)}$ optimization runs required by random initialization approaches.
We also propose a new parametrization of QAOA that may significantly simplify optimization by reducing the dimension of the search space. Using our heuristic strategy, we benchmark the performance of QAOA on many instances of MaxCut up to $N \le 22$ vertices and level $p\le 50$. Comparing QAOA with quantum annealing, we find the former can learn via optimization to utilize diabatic mechanisms~\cite{Crosson2014, Muthukrishnan2016Tunneling, Hormozi2017, Albash2018Adiabatic} and overcome the challenges faced by adiabatic quantum annealing due to very small spectral gaps.
Considering realistic experimental implementations, we also study the effects of quantum ``projection noise'' in measurement:
we find that, for numerically accessible problem sizes, QAOA can often obtain the solution among measurement outputs before the best variational parameters are found. 
Parameter optimization will be more useful at large system sizes (a few hundred vertices), as one expects the probability of finding the solution from projective measurements to decrease exponentially.
At such system sizes, we analyze a procedure to make graphs more experimentally realizable by reducing the required range of qubit interactions via vertex renumbering.
Finally, we discuss a specific implementation using neutral atoms interacting via Rydberg excitations~\cite{Bernien2017Probing, Saffman2010Quantum}, where a 2D implementation with a few hundred atoms appears feasible on a near-term device. 

The rest of the paper is organized in the following order: In Sec.~\ref{sec:background}, we review QAOA and the MaxCut problem. In Sec.~\ref{Sec:Heuristics}, we describe some patterns found for QAOA optimal parameters and introduce heuristic optimization strategies based on the observed patterns. We benchmark our heuristic strategies and study the performance of QAOA on typical MaxCut graph instances in Sec.~\ref{Sec:Performance}. We then, in Sec.~\ref{Sec:QAOAvsQA}, compare QAOA with quantum annealing, shedding light on the non-adiabatic mechanism of QAOA. Lastly, we discuss considerations for experimental implementations for large problem sizes in Sec.~\ref{Sec:Experiment}.


\begin{figure}[t]
\includegraphics[width=\linewidth]{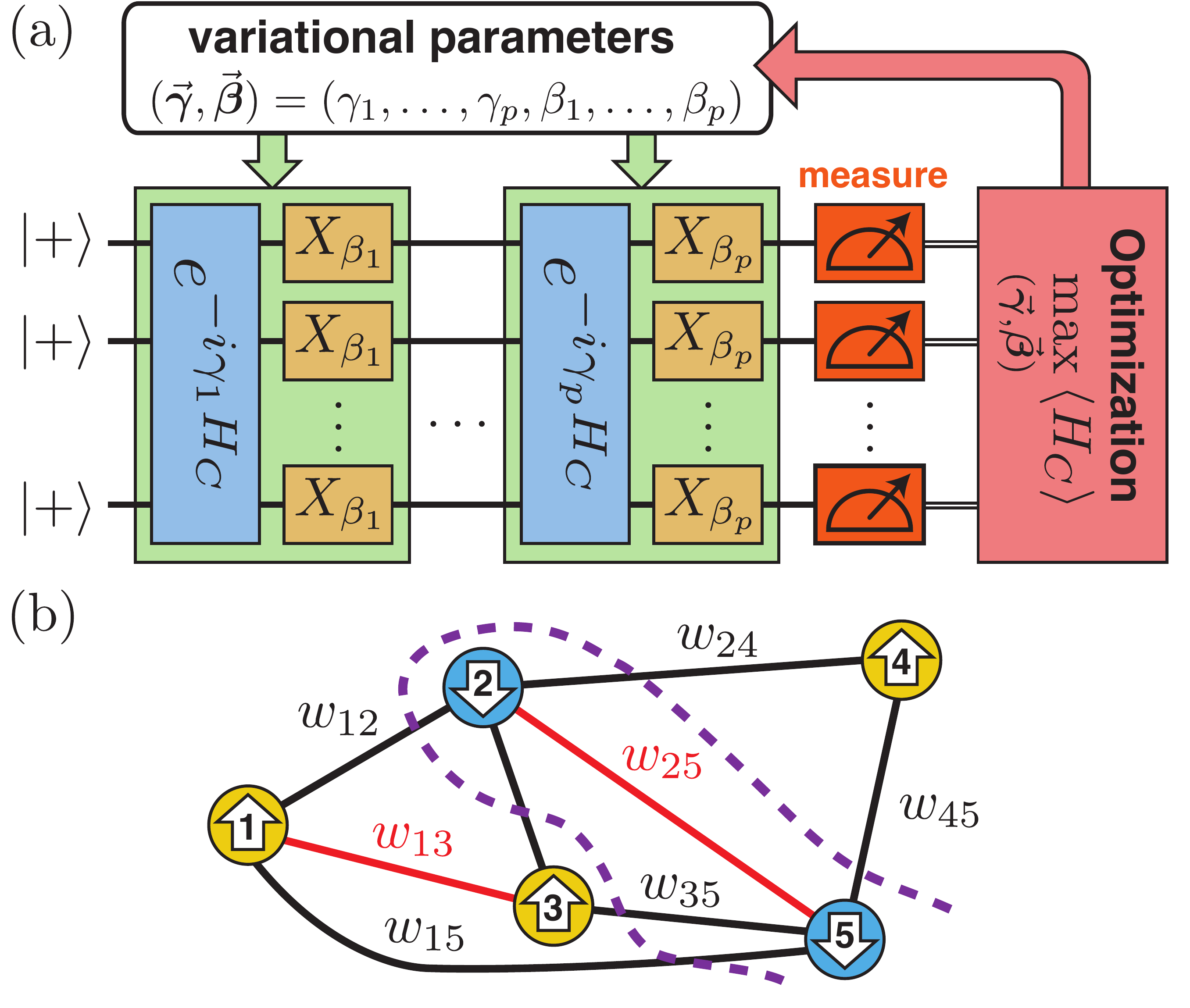}
\caption{\label{fig:intro}
(a) Schematic of a $p$-level Quantum Approximation Optimization Algorithm~\cite{QAOA}.
A quantum circuit takes input $\ket{+}^{\otimes n}$ and alternately applies $e^{-i\gamma_i H_C}$ and $X_{\beta_i} = e^{-i\beta_i \sigma^x}$, and the final state is measured to obtain expectation value with respect to the objective function $H_C$. This is fed to a (classical) optimizer to find the best parameters $(\vgamma,\vbeta)$ that maximizes $\ave{H_C}$.
(b) An example of a MaxCut problem on a 5-vertex graph, where one seeks 
an assignment of spin variables on the vertices such that the sum of edge weights between anti-aligned spins is maximized (black edges). 
}
\end{figure}

\section{Quantum Approximate Optimization Algorithm \label{sec:background}}
Many interesting real-world problems can be framed as combinatorial optimization problems~\cite{Papadimitriou1998Combinatorial, Korte2012Combinatorial}.
These are problems defined on $N$-bit binary strings $\vect{z}=z_1\cdots z_N$, where the goal
is to determine a string that maximizes a given classical objective function
$C(\vect{z}):\{+1,-1\}^N \mapsto\mathds{R}_{\ge0}$.
An approximate optimization algorithm aims to find a string $\vect{z}$ that achieves a desired approximation ratio
\begin{equation}
\frac{C(\vect{z})}{C_\text{max}} \ge r^*,
\end{equation}
where $C_\text{max}=\max_{\vect{z}}C(\vect{z})$.

The Quantum Approximate Optimization Algorithm (QAOA) is a quantum algorithm recently introduced to tackle these combinatorial optimization problems~\cite{QAOA}.
To encode the problem, the classical objective function can be converted to a quantum problem Hamiltonian by promoting each binary variable $z_i$ to a quantum spin $\sigma^z_i$:
\begin{equation}
H_C = C(\sigma^z_1, \sigma_2^z,\cdots, \sigma_N^z).
\end{equation}
For $p$-level QAOA, which is visualized in Fig.~\ref{fig:intro}(a), we initialize the quantum processor in the state $\ket{+}^{\otimes N}$, and then apply the problem Hamiltonian $H_{C}$ and a mixing Hamiltonian $H_{B} = \sum_{j=1}^N \sigma_j^x$ alternately with controlled durations to generate a variational wavefunction
\begin{equation}
\ket{\psi_{p}(\vgamma, \vbeta)} = e^{-i \beta_{p} H_{B}} e^{-i \gamma_{p} H_{C}} \cdots e^{-i \beta_{1} H_{B}} e^{-i \gamma_{1} H_{C}}  \ket{+}^{\otimes N},
\end{equation}
which is parameterized by $2p$ variational parameters $\gamma_{i}$ and $\beta_{i}$ ($i = 1,2,\cdots p$).
We then determine the expectation value $H_{C}$ in this variational state
\begin{equation}
F_{p}(\vgamma, \vbeta) = \bracket{\psi_{p}(\vgamma, \vbeta)}{H_{C}}{\psi_{p}(\vgamma, \vbeta)},
\end{equation}
which is done by repeated measurements of the quantum system in the computational basis.
A classical computer is used to search for the optimal parameters $(\vgamma^{*}, \vbeta^{*})$ so as to maximize the averaged measurement output $F_{p}(\vgamma, \vbeta)$,
  \begin{equation}
(\vgamma^{*}, \vbeta^{*}) = \arg \max_{\vgamma, \vbeta} F_{p}(\vgamma, \vbeta). 
\end{equation}
This is typically done by starting with some initial guess of the parameters and performing simplex or gradient-based optimization.
A figure of merit for benchmarking the performance of QAOA is the approximation ratio
\begin{equation}
r = \dfrac{ F_{p}(\vgamma^{*}, \vbeta^{*}) }{C_{\max}}.
\end{equation}

The framework of QAOA can be applied to general combinatorial optimization problems. Here, we focus on its application to an archetypical problem called MaxCut, which is a combinatorial problem whose approximate optimization beyond a minimum ratio $r^*$ is NP-hard~\cite{Hastad2001, Berman1999}.
The MaxCut problem, visualized in Fig.~\ref{fig:intro}(b), is defined for any input graph $G=(V,E)$.
Here, $V=\{1,2,\ldots,N\}$ denotes the set of vertices and $E=\{(\ave{i,j},w_{ij})\}$ is the set of edges, where $w_{ij}\in \mathds{R}_{\ge0}$ is the weight of the edge $\ave{i,j}$ connecting vertices $i$ and $j$.
The goal of MaxCut is to maximize the following objective function
\begin{equation}
H_{C} = \sum_{\left<i,j \right>} \frac{w_{i j}}{2} (\Id - \sigma_{i}^{z} \sigma_{j}^{z}),
\end{equation}
where an edge $\ave{i,j}$ contributes with weight $w_{ij}$ if and only if spins $\sigma_i^z$ and $\sigma_j^z$ are anti-aligned.

For simplicity, we restrict our attention to MaxCut on $d$-regular graphs, where every vertex is connected to exactly $d$ other vertices.
We study two classes of graphs:
the first is unweighted $d$-regular graphs (u$d$R), where all edges have equal weights $w_{ij}=1$; the second is weighted $d$-regular graphs (w$d$R), where the weights $w_{ij}$ are chosen uniformly at random from the interval $[0,1]$.
It is NP-hard to design an algorithm that guarantees a minimum approximation ratio of $r^*\ge 16/17$ for MaxCut on all graphs~\cite{Hastad2001}, or $r^*\ge331/332$ when restricted to u3R graphs~\cite{Berman1999}.
The current record for approximation ratio guarantee 
on generic graphs belongs to Goemans-Williamson~\cite{GW}, which achieves $r^*\approx 0.87856$ using semi-definite programming. This lower bound can be raised to $r^*\approx 0.9326$ when restricted to u3R graphs~\cite{Halperin2004}. 
Farhi \etal~\cite{QAOA} were able to prove that QAOA at level $p=1$ achieves $r^*\ge 0.6924$ for u3R graphs, using the fact that $F_p$ can be written as a sum of quasi-local terms, each corresponding to a subgraph involving edges at most $p$ steps away from a given edge.
However, this approach to bound $r^*$ quickly becomes intractable since the locality of each term (i.e., size of each subgraph) grows exponentially in $p$, as does the number of subgraph types involved.

QAOA is believed to be a promising algorithm for multiple reasons~\cite{QAOA, QAOAE3lin, QAOAfermionic, QAOAsupremacy, Wecker2016Training, YangVQA, Ho2018Efficient, QAOAGrover, Anschuetz2018Variational}.
As mentioned above, for certain cases one can prove a guaranteed minimum approximation ratio when $p=1$~\cite{QAOA, QAOAE3lin}.
Additionally, under reasonable complexity-theoretic assumptions, QAOA cannot be efficiently simulated by any classical computer even when $p =1$, making it a suitable candidate algorithm to establish the so-called ``quantum supremacy''~\cite{QAOAsupremacy}. 
It has also been argued that the square-pulse (``bang-bang'') ansatz of dynamical evolution, of which QAOA is an example, is optimal given a fixed quantum computation time~\cite{YangVQA}.
In general, the performance of QAOA can only improve with increasing $p$, achieving $r\to 1$ when $p\to\infty$ since it can approximate adiabatic quantum annealing via Trotterization; this monotonicity makes it more attractive than quantum annealing whose performance may decrease with increased run time~\cite{Crosson2014}.

While QAOA has a simple description, not much is currently understood beyond $p =1$. To establish potential quantum advantage over classical algorithms, it is of critical importance to investigate QAOA at intermediate depths ($p>1$).
Refs.~\cite{QAOA, Hastings2019Classical, BKKT2019} have shown that QAOA have limited performance on some problems on bounded-degree graphs when the depth is shallow.
This limitation may result from the fact that the algorithm cannot ``see'' the entire graph at low depth.
It thus indicates that one may need the depth of QAOA to grow with the system size (e.g., $p\ge \log N$) in order to outperform the best classical algorithms.
For the toy example of MaxCut on u2R graphs, i.e.\ 1D antiferromagnetic rings, it is conjectured that QAOA yields $r \ge (2p+1)/(2p+2)$ based on numerical evidence \cite{QAOA, QAOAfermionic}\footnote{{The approximation ratio $r = (2p+1)/(2p+2)$ is found for an infinite ring. For a finite ring with $N$ vertices, numerical calculations show that for $p <  \left \lfloor N/2 \right \rfloor$ one has $r = (2p+1)/(2p+2)$ for even $N$ and $r = \frac{(2p+1)(N+1)}{(2p+2)N}$ for odd $N$, and for $p \geq \left \lfloor N/2 \right \rfloor$ one has $r = 1$}}.
In another example of Grover's unstructured search problem among $n$ items, QAOA is shown to be able to find the target state with $p=\Theta(\sqrt{n})$, achieving the optimal query complexity within a constant factor~\cite{QAOAGrover}. 
For more general problems, Farhi \etal~\cite{QAOA} proposed a simple approach by discretizing each parameter into $O(\poly(N))$ grid points; this, however, requires examining $N^{O(p)}$ possibilities at level $p$, which quickly becomes impractical as $p$ grows.
Efficient optimization of QAOA parameters and understanding the algorithm for $1< p< \infty$ remain outstanding problems.
We address these problems in the present work.

\begin{figure}[t]
\includegraphics[width=\linewidth]{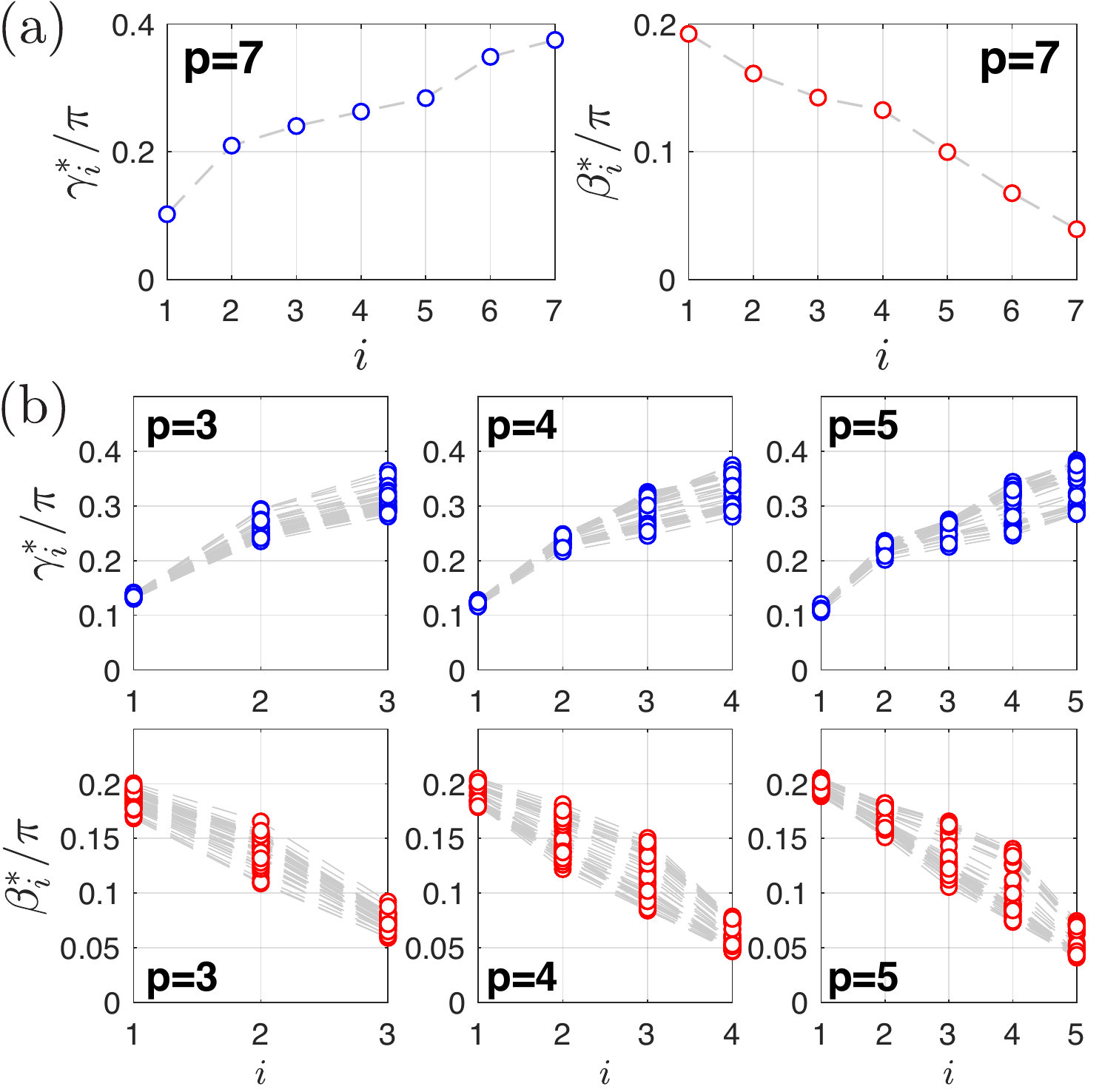}

\caption{\label{fig:param-pattern}
(a) Optimal QAOA parameters $(\vgamma^*, \vbeta^*)$ for an example instance of MaxCut on a 16-vertex unweighted 3-regular (u3R) graph at level $p=7$.
(b) The parameter pattern visualized by plotting the optimal parameters of 40 instances of $16$-vertex u3R graphs, for $3\le p\le 5$. Each dashed line connects parameters for one particular graph instance. For each instance and each $p$, we use the classical BFGS optimization routine~\cite{BFGS1, *BFGS2, *BFGS3, *BFGS4} from $10^4$ random initial points, and keep the best parameters.
}
\end{figure}


\vspace{-5pt}
\section{Optimizing Variational Parameters}
\label{Sec:Heuristics}

\vspace{-5pt}

In this section, we address the issue of parameter optimization in QAOA, since searching for the best parameters via standard approaches that rely on random initialization generally become exponentially difficult as level $p$ increases.
%
We mostly restrict our discussion to randomly generated instances of u3R and w3R graphs.
Similar results are found for u4R and w4R graphs, as well as complete graphs with random weights. 
We utilize patterns in the optimal parameters to develop heuristic strategies that can efficiently find quasi-optimal solutions in $O(\poly(p))$ time.

\subsection{Patterns in optimal parameters}

Before searching for patterns in optimal QAOA parameters, 
it is useful to eliminate degeneracies in the parameter space due to symmetries.
Generally, QAOA has a time-reversal symmetry, $F_p(\vgamma,\vbeta)=F_p(-\vgamma,-\vbeta)$, since both $H_B$ and $H_C$ are real-valued.
For QAOA applied to MaxCut, there is an additional $\mathds{Z}_2$ symmetry, as $e^{-i (\pi/2) H_{B}}\equiv(\sigma^x)^{\otimes N}$ commutes through the circuit.
Furthermore, the structure of the MaxCut problem on u$d$R graphs creates redundancy since $e^{-i\pi H_C}=\Id$ if $d$ is even, and $(\sigma^z)^{\otimes N}$ if $d$ is odd.
These symmetries allow us to restrict $\beta_i \in[-\frac{\pi}{4}, \frac{\pi}{4})$ in general, and
$\gamma_i \in [-\frac{\pi}{2},\frac{\pi}{2})$ for u$d$R graphs.

We start by numerically investigating the optimal QAOA parameters for MaxCut on random u3R and w3R graphs with vertex number $8\le N \le 22$, with extensive searches in the entire parameter space.
For each graph instance and a given level $p$, we choose a random initial point (seed) in the parameter space~\footnote{The initial points $\beta^0_i$ are drawn uniformly from $[-\frac{\pi}{4}, \frac{\pi}{4})$, and $\gamma^0_i$ are drawn uniformly $[-\frac{\pi}{2}, \frac{\pi}{2})$ for u3R graphs or $[-2\pi, 2\pi)$ for w3R graphs. Although $\gamma_i^0$ can meaningfully take values beyond the restricted range $\gamma_i^0\in[-2\pi,2\pi)$ for w3R graph, we find that broadening the range does not improve performance. The ranges of the output parameters are not restricted in our unconstrained optimization routine.} and use a commonly used, gradient-based optimization routine known as BFGS~\cite{BFGS1,*BFGS2,*BFGS3,*BFGS4} to find a local optimum $(\vgamma^L,\vbeta^L)$.
This local optimization is repeated with sufficiently many different seeds to find the global optimum $(\vgamma^*,\vbeta^*)$~\footnote{We denote the best of all local optima optimized from $k$ random initial points (seeds) to be $(\vgamma^{B[k]},\vbeta^{B[k]})$. When the same optimum $(\vgamma^{B[k]},\vbeta^{B[k]})$ is found from many different seeds and continues to yield the best $F_p(\vgamma^{B[k]},\vbeta^{B[k]})$ as we increase the number of seeds $k$, we then claim that it is a global optimum, i.e., $(\vgamma^*,\vbeta^*)=\lim_{k\to\infty} (\vgamma^{B[k]},\vbeta^{B[k]})$.}.
We then reduce the degeneracies of the optimal parameters $(\vgamma^*,\vbeta^*)$ using the aforementioned symmetries. In all cases examined, we find that the global optimum is non-degenerate up to these symmetries.

After performing the above numerical experiment for 100 random u3R and w3R graphs with various vertex number $N$, we discover some patterns in the optimal parameters $(\vgamma^*,\vbeta^*)$.
Generically, the optimal $\gamma^*_i$ tends to increase smoothly with $i=1,2,\cdots\!,p$, while $\beta^*_i$ tends to decrease smoothly, as shown for the example instance in Fig.~\ref{fig:param-pattern}(a).
In Fig.~\ref{fig:param-pattern}(b), we illustrate the pattern by simultaneously plotting the optimal parameters for 40 instances of 16-vertex u3R graphs for $3\le p \le 5$.
Furthermore, for a given class of graphs, the optimal parameters are observed to roughly occupy the same range of values as $p$ is varied.
Similar patterns are found for w3R graphs and weighted complete graphs, which we illustrate in Appendix~\ref{appx:param-pattern}.
This demonstrates a clear pattern in the optimal QAOA parameters that we can exploit in the optimization, as we discuss later in Sec.~\ref{Sec:heuristic}.
Similar patterns are found for parameters up to $p\lesssim 15$, if the number of random seeds is increased accordingly.

We give two remarks here:
First, we note that surprisingly, even at small depth, this parameter pattern is reminiscent of adiabatic quantum annealing where $H_C$ is gradually turned on while $H_B$ is gradually turned off.
However, we will demonstrate in Sec.~\ref{Sec:QAOAvsQA} that the mechanism of QAOA goes beyond the adiabatic principle.
Secondly, we note that these optimal parameters have a small spread over many different instances.
This is because the objective function $F_p(\vgamma, \vbeta)$ is a sum of terms corresponding to subgraphs involving vertices that are distance $\le p$ away from every edge.
At small $p$, there are only a few relevant subgraph types that enter into $F_p$ and effectively determine the optimal parameters.
As $N\to\infty$ and at a fixed finite $p$, we expect the probability of a relevant subgraph type appearing in a random graph to approach a fixed fraction.
This implies that the distribution of optimal parameters $(\vgamma^*,\vbeta^*)$ should converge to a fixed set of values in this limit.

\subsection{Heuristic optimization strategy for large $p$\label{Sec:heuristic}}
 
The optimal parameter patterns observed above indicate that generically, there is a slowly varying continuous curve that underlies the parameters $\gamma^*_i$ and $\beta^*_i$.
Moreover, this curve changes only slightly from level $p$ to $p+1$.
These observations allow us to choose educated guesses of variational parameters for $(p+1)$-level QAOA based on optimized parameters from $p$-level (or in general from $q$-level, where $q\leq p$).
These educated guesses can serve as initial points fed to various classical optimization routines that find a nearby local optimum.
Based on this idea,
we have developed two types of heuristic strategies for initializing optimization.
%
%
%
We give a high-level overview of these heuristics in this section, while deferring the details of its implementation to Appendix~\ref{appx:heuristics}.
While these heuristics are not guaranteed to find the global optimum of QAOA parameters, we show that in Sec.~\ref{sec:comp-BF}, it can produce, in $O(\poly(p))$ time,  quasi-optima that require $2^{O(p)}$ randomly initialized optimization runs to surpass.
Consequently, this allows us to study the performance and mechanism of QAOA beyond $p = 1$.

The first heuristic strategy, which we call INTERP, uses linear interpolation to choose initial parameters.
Starting at level $p$\,=\,1, we optimize, and linearly interpolate the curve formed by optimized parameters at level $p$ to extract a set of initial parameters for level $p+1$.

The second heuristic strategy, which we call FOURIER, uses a new parameterization of QAOA.
%
Instead of using the $2p$ parameters $(\vgamma,\vbeta)\in\mathds{R}^{2p}$, we switch to $2q$ parameters $(\vu,\vv)\in\mathds{R}^{2q}$, where the individual elements $\gamma_i$ and $\beta_i$ are written as functions of $(\vu,\vv)$ through the following transformation:
\begin{equation} \label{eq:uv-rep}
\begin{split}
\gamma_i &= \sum_{k=1}^q u_k \sin\left[\left(k-\frac12 \right)\left(i-\frac12 \right)\frac{\pi}{p}\right], \\
\beta_i  &= \sum_{k=1}^q v_k \cos\left[\left(k-\frac12\right)\left(i-\frac12 \right)\frac{\pi}{p}\right].\
\end{split}
\end{equation}
These transformations are known as Discrete Sine/Cosine Transform, where $u_k$ and $v_k$ can be interpreted as the amplitude of $k$-th frequency component for $\vgamma$ and $\vbeta$, respectively.
In this strategy, when optimizing level $p+1$, the initial parameters  are generated by simply re-using the optimized amplitudes $(\vu^*,\vv^*)$ from level $p$.
Note that when $q\ge p$, the $(\vu,\vv)$ parametrization is capable of describing all possible QAOA protocols at level $p$.
However, the smoothness of the optimal parameters $(\vgamma, \vbeta)$ implies that only the low-frequency components are important. Thus, we can also consider the case where $q$ is a fixed constant independent of $p$, so the number of parameters is bounded even as the QAOA circuit depth increases.

While both heuristics work well, we have decided to focus on using the FOURIER heuristics in the main text, because we find that it gives a slight edge in performance for MaxCut problems. The INTERP heuristic is simpler and may work better on other problems.
More details on the implementation of the heuristics and their comparison can be found in Appendix~\ref{appx:heuristics}.

We stress that these heuristic strategies are developed to generate good initial points for optimization.
These initial points can then be fed to standard optimization routines such as gradient descent, BFGS~\cite{BFGS1, *BFGS2, *BFGS3, *BFGS4}, Nelder-Mead~\cite{NelderMead}, and Bayesian Optimization~\cite{bayesopt}.
This is in contrast to the standard strategy of random initialization (RI), where one picks a random set of parameters to begin optimization.
In order to find the global optimum in a highly non-convex landscape,
the number of RI runs needed generically scales exponentially with the number of parameters, which becomes intractable for large number of parameters.
In the following section, we will compare our heuristics to the RI approach, and find that an exponential number of RI runs is needed to match the performance of our heuristics.


\begin{figure}[t!]
\includegraphics[width=\linewidth]{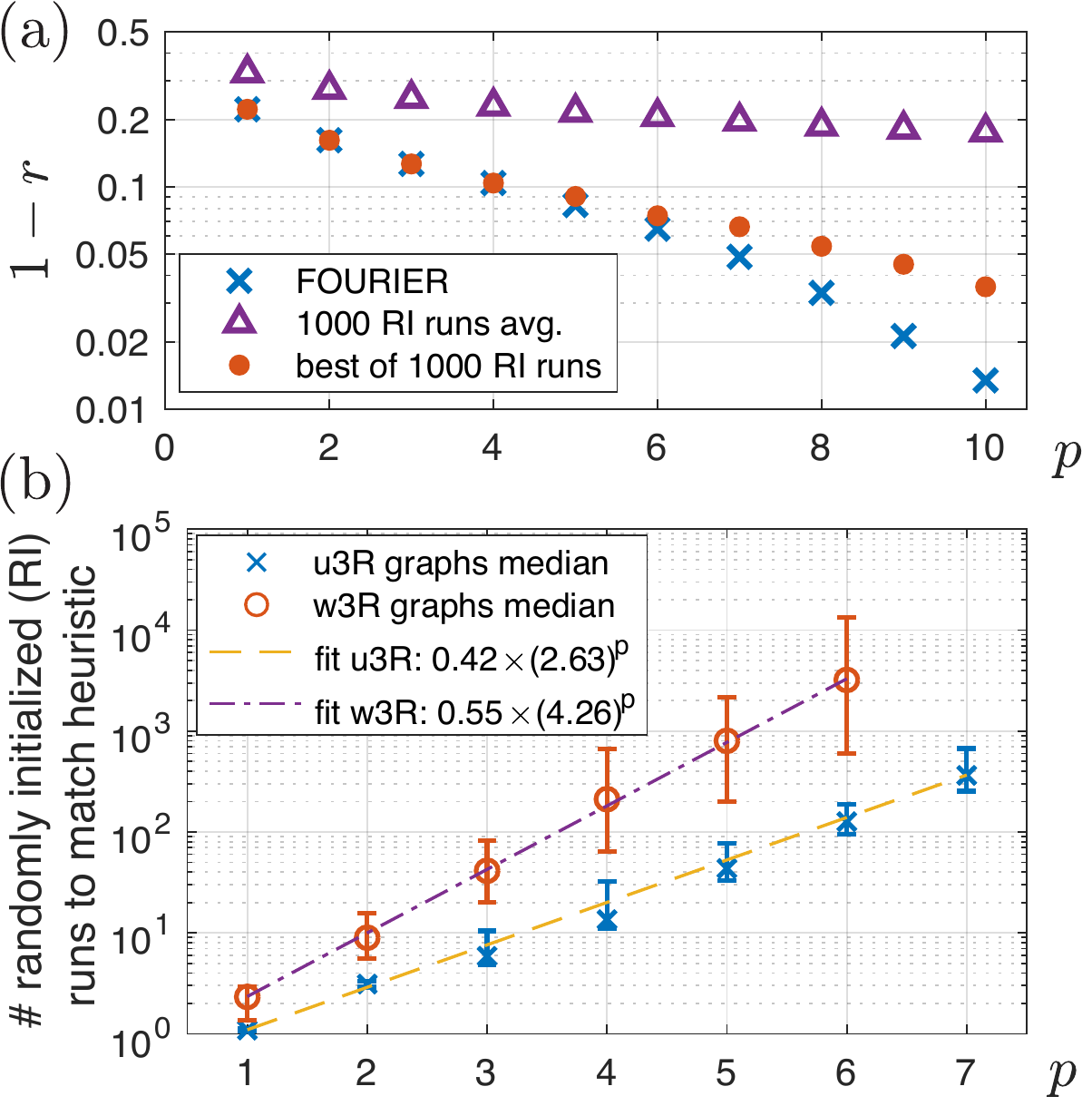}
\caption{\label{fig:heuristic}
(a) Comparison between our FOURIER heuristic and the random initialization (RI) approach for optimizing QAOA, on an example instance of 16-vertex w4R graph. The figure of merit $1-r$, where $r$ is the approximation ratio, is plotted as a function of QAOA level $p$ on a log-linear scale. The RI points are obtained by optimizing from 1000 randomly generated initial parameters, averaged over 10 such realizations.
(b) The median number of randomly initialized optimization runs needed to obtain an approximation ratio as good as our FOURIER heuristic, for 40 instances of 16-vertex u3R and w3R graphs. A log-linear scale is used, and exponential curves are fitted to the results. Error bars are 5th and 95th percentiles.
}
\end{figure}

\section{Performance of heuristically optimized QAOA}
\label{Sec:Performance}

\subsection{Comparison between our heuristics and randomly initialized (RI) optimizations \label{sec:comp-BF}}

Here, we compare the performance of our heuristic strategies to the standard strategy of random initialization (RI).
In Fig.~\ref{fig:heuristic}(a), we show the result for an example instance of 16-vertex w4R graph.
At low $p$, our FOURIER heuristic strategies perform just as well as the best out of 1000 RI optimization runs---both are able to find the global optimum. The average performance of the 
RI
strategy, on the other hand, is much worse than our heuristics.
This indicates that the QAOA parameter landscape is highly non-convex and filled with low-quality, non-degenerate local optima.
When $p\ge 5$, our heuristic optimization outperforms the best RI run.
To estimate the number of RI runs needed to find an optimum with the same or better approximation ratio as our FOURIER heuristics, we generate 40 instances of 16-vertex u3R and w3R graphs, and perform 40000 RI optimization for each instance at each level $p$.
In Fig.~\ref{fig:heuristic}(b), the median number of RI runs needed to match our heuristic is shown to scale exponentially with $p$.
Therefore, our heuristics offer a dramatic improvement in the resource required to find good parameters for QAOA.
As we have verified with an excessive number of RI runs, the heuristics usually find the global optima.

We remark that although we mostly used a gradient-based optimization routine (BFGS) in our numerical simulations, non-gradient based routines, such as Nelder-Mead~\cite{NelderMead}, work equally well with our heuristic strategies.
The choice to use BFGS is mainly motivated by the simulation speed.
Later in Sec.~\ref{Sec:Experiment}, we account for the measurement costs in estimating the gradient using a finite-difference method, and perform a full Monte-Carlo simulation of the entire QAOA algorithm, including quantum fluctuations in the determination of $F_p$.

\subsection{Performance of QAOA on typical instances\label{Sec:performance-typical}}

\begin{figure}[t]
\includegraphics[width=\linewidth]{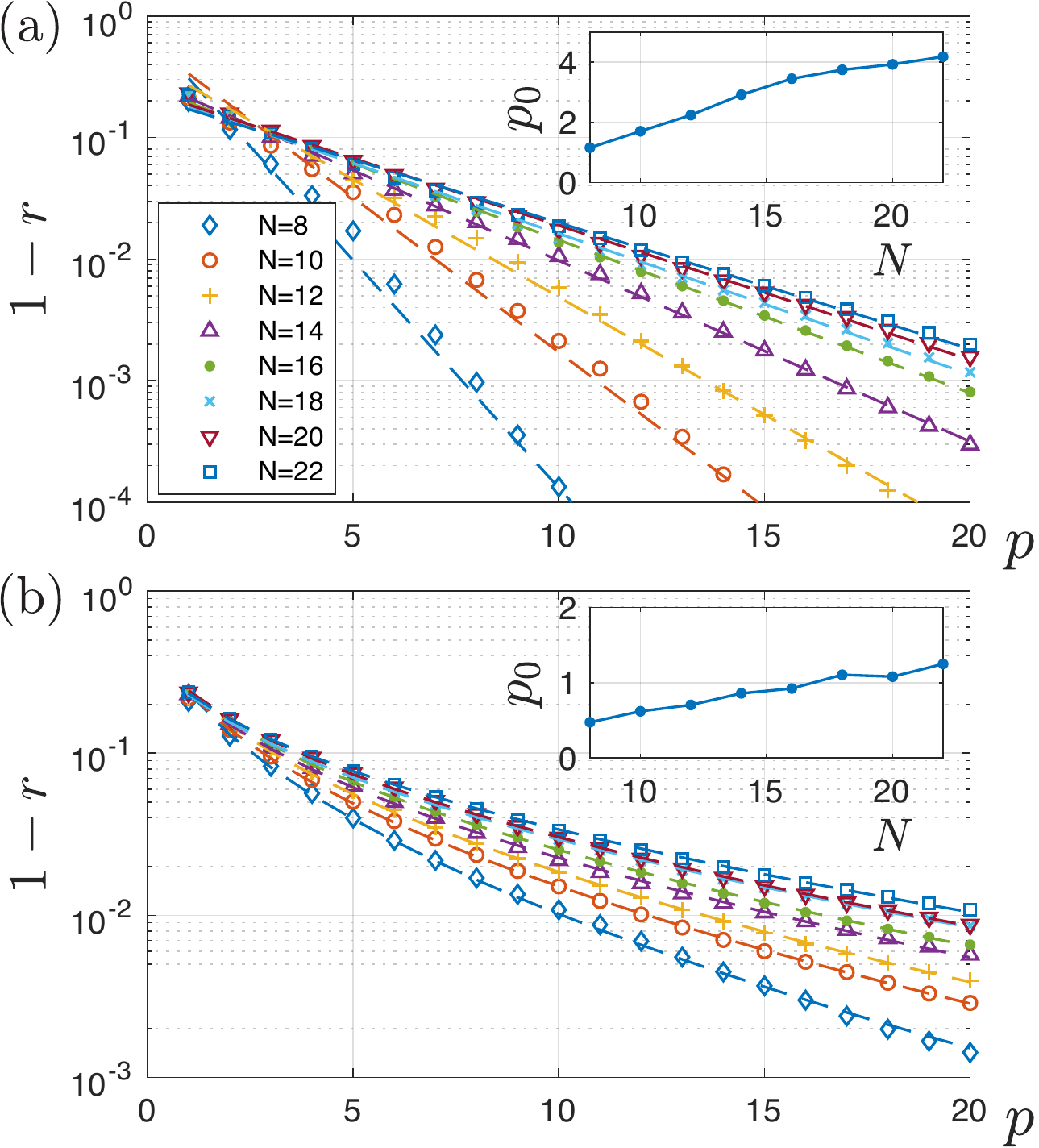}
\caption{\label{fig:performance}Average performance of QAOA as measured by the fractional error $1-r$, plotted on log-linear scale.
The results are obtained by applying our heuristic optimization strategies FOURIER to up to 100 random instances of (a) u3R graphs and (b) w3R graphs.
Differently colored lines correspond to fitted lines to the average for different system size $N$, where the model function is $1-r\propto e^{-p/p_0}$ for unweighted graphs and $1-r\propto e^{-\sqrt{p/p_0}}$ for weighted graphs.
Insets show the dependence of the fit parameters $p_{0}$ on the system size $N$.
}
\end{figure}

With our heuristic optimization strategies in hand, we study the performance of intermediate $p$-level QAOA on many graph instances.
We consider many randomly generated instances u3R and w3R graphs with vertex number $8\le N \le 22$, and use our FOURIER strategy to find optimal QAOA parameters for level $p\le 20$.
In the following discussion, we use the fractional error $1-r$ to assess the performance of QAOA.

We first examine the case of unweighted graphs, specifically u3R graphs.
In Fig.~\ref{fig:performance}(a), we plot the fractional error $1-r$ as a function of QAOA's level $p$.
Here, we see that, on average, $1- r \propto e^{-p/p_0}$ appears to decay exponentially with $p$.
Note that since the instances studied are u3R graphs with system size $N \le 22$, where $C_\text{max} \le |E| \le 33$, we have essentially prepared the MaxCut state whenever the fractional error goes below $\sim10^{-2}$.
This good performance can be understood by interpreting QAOA as Trotterized quantum annealing, especially when the optimized parameters are of the pattern seen in Fig.~\ref{fig:param-pattern}, where one initializes in the ground state of $-H_B$ and evolve with $H_B$ (and $H_C$) with smoothly decreasing (and increasing) durations.
The equivalent total annealing time $T$ is approximately proportional to the level $p$, since the individual parameters $\gamma_i, \beta_i = O(1)$ and correspond to the evolution times under $H_C$ and $H_B$.
If $T$ is much longer than $1/\Delta_{\min}^2$, where $\Delta_{\min}$ is the minimum spectral gap, quantum annealing will be able to find the exact solution to MaxCut (ground state of $-H_C$) by the adiabatic theorem~\cite{Albash2018Adiabatic},
and achieve exponentially small fractional error as predicted by Landau-Zener~\cite{Landau1932, *Zener1932}.
Numerically, we find that the minimum gaps of these u3R instances when running quantum annealing are on the order of $\Delta_{\min}\gtrsim0.2$.
It is thus not surprising that QAOA achieves exponentially small fractional error on average, since it is able to prepare the ground state of $-H_C$ through the adiabatic mechanism for these large-gap instances.
Nevertheless, as we will see in the following section, this exponential behavior breaks down for hard instances, where the gap is small.

Let us now turn to the case of w3R graphs.
As shown in Fig.~\ref{fig:performance}(b), the fractional error appears to scale as $1-r \propto e^{-\sqrt{p/p_0}}$.
We note that the stretched-exponential scaling is true in the average sense, while individual instances have very different behavior.
For easy instances whose corresponding minimum gaps $\Delta_{\min}$ are large, exponential scaling of the fractional error is found.
For more difficult instances whose minimum gaps are small, we find that their fractional errors reach plateaus at intermediate $p$, before decreasing further when $p$ is increased.
We analyze these hard instances in more depth in the following Section~\ref{Sec:QAOAvsQA}.
Surprisingly, when we average over randomly generated instances, the fractional error is fitted remarkably well by a stretched-exponential function.

These results of average performance of QAOA are interesting despite considerations of finite-size effect.
While the decay constant $p_0$ does appear to depend on the system size $N$ as shown in the insets of Fig.~\ref{fig:performance}, our finite-size simulations cannot conclusively determine the exact scaling.
Although it remains unknown whether the (stretched)-exponential scaling will start to break down at larger system sizes, if the trend continues to system size of $N = 100\sim 1000$, then QAOA will be practically useful in solving interesting MaxCut instances, better or on par with other known algorithms, in a regime where finding the exact solution will be infeasible.
Even if QAOA fails for the worst-case graphs, it can still be practically useful, if it performs well on a randomly chosen graph of large system size.

\section{Adiabatic mechanism, quantum annealing, and QAOA}
\label{Sec:QAOAvsQA}

In the previous section, we benchmark the performance of QAOA for MaxCut on random graph instances in terms of the approximation ratio $r$. Although designed for approximate optimization, QAOA is often able to find the MaxCut configuration---the global optimum of the problem---with a high probability as level $p$ increases.
In this section, we assess the efficiency of the algorithm to find the MaxCut configuration and compare it with quantum annealing.
In particular, we find that QAOA is not necessarily limited by the minimum gap as in quantum annealing, and explain a mechanism at work that allows it to overcome the adiabatic limitations.   

\subsection{Comparing QAOA with quantum annealing}

A predecessor of QAOA, quantum annealing (QA) has been widely studied for the purpose of solving combinatorial optimization problems \cite{Kadowaki98Quantum, Farhi2001Quantum, Boixo2014Evidence, Ronnow2014Defining}.
To find the \mbox{MaxCut} configuration that maximizes $\ave{H_C}$, we consider the following simple QA protocol:
\begin{equation} \label{Eq:HamQA}
H_{\text{QA}}(s) = -[s H_{C} + (1-s) H_{B}], \qquad s = t/T, 
\end{equation}
where $t \in [0,T]$ and $T$ is the total annealing time.
The initial state is prepared to be the ground state of $H_{\text{QA}}(s=0)$, i.e., $\ket{\psi(0)}=\ket{+}^{\otimes N}$.
The ground state of the final Hamiltonian, $ H_{\text{QA}}(s=1)$, corresponds to the solution of the MaxCut problem encoded in $H_{C}$
\footnote{To be consistent with the language of QA, here we use the terminology of ground state and low excited states of $-H_{C}$ instead of referring to them as highest excited states in the MaxCut language.}%
.
In adiabatic QA, the algorithm relies on the adiabatic theorem to remain in the instantaneous ground state along the annealing path, and solves the computational problem by finding the ground state at the end.
To guarantee success, the necessary run time of the algorithm typically scales as $T = O(1/\Delta_{\min}^{2})$, where $\Delta_{\min}$ is the minimum spectral gap~\cite{Albash2018Adiabatic}.
Consequently, adiabatic QA becomes inefficient for instances where $\Delta_{\min}$ is extremely small. We refer to these graph instances as hard instances (for adiabatic QA).
  
Beyond the completely adiabatic regime, there is often a tradeoff between the success probability (ground state population $p_{\text{GS}}(T)$) and the run time (annealing time $T$): one can either run the algorithm with a long annealing time to obtain a high success probability or run it multiple times at a shorter time to find the solution at least once. A metric often used to determine the best balance is the time-to-solution (TTS) \cite{Albash2018Adiabatic}: 
\begin{align}
\text{TTS}_{\text{QA}}(T) &= T \dfrac{\ln (1-p_{d})}{\ln [1-p_{\text{GS}}(T)]} \\
\text{TTS}_{\text{QA}}^{\text{opt}}  & = \min_{T>0} \text{TTS}_{\text{QA}}(T). 
\end{align}
$\text{TTS}_{\text{QA}}(T)$ measures the time required to find the ground state at least once with the target probability $p_{d}$ (taken to be $99\%$ in this paper), neglecting non-algorithmic overheads such as state-preparation and measurement time.
In the adiabatic regime where 
$\ln[1-p_{\text{GS}}(T)] \propto T \Delta_{\min}^{2}$ 
per Landau-Zener formula~\cite{Landau1932, *Zener1932}, we have $\text{TTS}_{\text{QA}}\propto 1/\Delta_{\min}^2$ which is independent of $T$.
However, it has been observed in some cases that it can be better to run QA non-adiabatically to obtain a shorter TTS~\cite{Crosson2014, Muthukrishnan2016Tunneling, Hormozi2017, Albash2018Adiabatic}.
By choosing the best annealing time $T$, regardless of adiabaticity, we can determine $\text{TTS}_{\text{QA}}^{\text{opt}}$ as the minimum algorithmic run time of QA.
For QAOA, a similar metric can be defined.
The variational parameters $\gamma_{i}$ and $\beta_{i}$ can be regarded as the time evolved under the Hamiltonians $H_{C}$ and $H_{B}$, respectively.
One can thus interpret the sum of the variational parameters to be the total ``annealing'' time, i.e., $T_p = \sum_{i=1}^{p} (|\gamma_{i}^{*}| + |\beta_{i}^{*}|)$ \footnote{This could change with different normalizations of the Hamiltonians $H_{C}$ and $H_{B}$, but our qualitative results remain the same. The physical limitation in the experiment is typically the interaction strength in $H_C$.}, and define 
\begin{align}
\text{TTS}_{\text{QAOA}}(p) &= T_p \dfrac{\ln (1-p_{d})}{\ln [1-p_{\text{GS}}(p)]} \\
\text{TTS}_{\text{QAOA}}^{\text{opt}}  & = \min_{p>0} \text{TTS}_{\text{QAOA}}(p), 
\end{align}
where $p_{\text{GS}}(p)$ is the ground state population after the optimal $p$-level QAOA protocol.
Note this quantity does not take into account of the overhead in finding the optimal parameters.
We use $\text{TTS}_{\text{QAOA}}(p)$ here to benchmark the algorithm, and it should not be taken directly to be the actual experimental run time.

\begin{figure}[t]
\includegraphics[width=\linewidth]{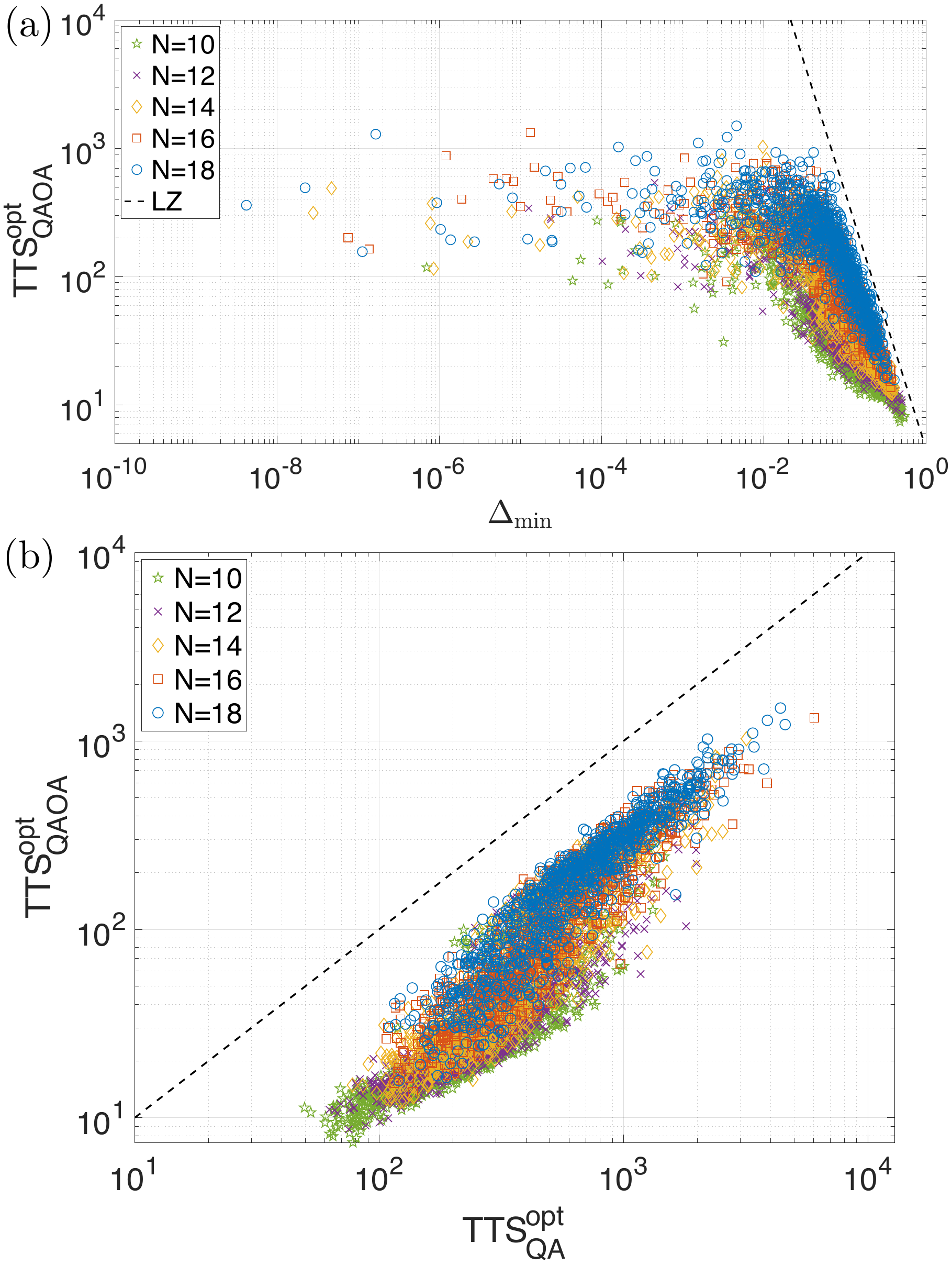} 
\caption{
(a) $\text{TTS}_{\text{QAOA}}^{\text{opt}}$ versus the minimum spectral gap in QA , $\Delta_{\min}$, for many random w3R graph instances. 1000 random instances are generated for each graph vertex size $N$. The maximum cutoff $p$ is taken to be $p_{\max} = 50,50,40,35,30$ for $N = 10,12,14,16,18$, respectively.
Dashed line corresponds to adiabatic QA run time of $1/\Delta_{\min}^2$ predicted by Landau-Zener~\cite{Landau1932, *Zener1932}.
(b) $\text{TTS}_{\text{QAOA}}^{\text{opt}}$ versus $\text{TTS}_{\text{QA}}^{\text{opt}}$ for each random graph instance.
Dashed line indicates when the two are equal. The (Pearson's) correlation coefficient between QAOA TTS and QA TTS is $\rho \left(\ln(\text{TTS}_{\text{QAOA}}^{\text{opt}}), \ln(\text{TTS}_{\text{QA}}^{\text{opt}}) \right) \approx 0.91$.}
\label{Fig:TTS}
\end{figure} 

To compare the algorithms, we compute $\text{TTS}_{\text{QA}}^{\text{opt}}$ and $\text{TTS}_{\text{QAOA}}^{\text{opt}}$ for many random graph instances.
For each even vertex number from $N=10$ to $N=18$, we randomly generate $1000$ instances of w3R graphs.
In Fig.~\ref{Fig:TTS}(a), we plot the relationship between $\text{TTS}_{\text{QAOA}}^{\text{opt}}$ and the minimum gap $\Delta_{\min}$ in quantum annealing for each instance. Most of the random graphs have large gaps ($\Delta_{\min} \gtrsim 10^{-2}$), and we observe that the optimal TTS follows the Landau-Zener prediction of $1/\Delta_{\min}^2$ for these graphs.
This indicates that a quasi-adiabatic parametrization of QAOA is the best when $\Delta_{\min}$ is reasonably large.
Many graphs, however, exhibit very small gaps ($\Delta_{\min} \lesssim 10^{-3}$), and thus require exceedingly long run time for adiabatic evolution.
For some graphs, $\Delta_{\min}$ is as small as $10^{-8}$, which implies an adiabatic evolution requires a run time $T \gtrsim 10^{16}$.
Nevertheless, we see that QAOA can find the solution for these hard instances faster than adiabatic QA.
Remarkably, $\text{TTS}_{\text{QAOA}}^{\text{opt}}$ appears to be independent of the gap for all graphs that have extremely small gaps, and beats the adiabatic TTS (Landau-Zener line) by many orders of magnitude.
This suggests that an exponential improvement of TTS is possible with non-adiabatic mechanisms when adiabatic QA is limited by exponentially small gaps.  

Similarly for QA, the optimal annealing time $T$ is often not in the adiabatic limit for small-gap graphs. 
In Fig.~\ref{Fig:TTS}(b), we observe a strong correlation between $\text{TTS}_{\text{QAOA}}^{\text{opt}}$ and $\text{TTS}_{\text{QA}}^{\text{opt}}$ for each graph instance.
This suggests that QAOA is making use of the optimal annealing schedule, regardless of whether a slow adiabatic evolution or a fast diabatic evolution is better.
We believe that if we use an optimized schedule beyond the linear ramp in Eq.~\eqref{Eq:HamQA}, QA should be able to match the performance of QAOA.
In the following subsection, we take a representative example to explain our results observed in Fig.~\ref{Fig:TTS} and a mechanism of QAOA to go beyond the adiabatic paradigm.

\subsection{Beyond the adiabatic mechanism: a case study}
\label{subSec:mechanism}

\begin{figure*}[t]
\includegraphics[width=\textwidth]{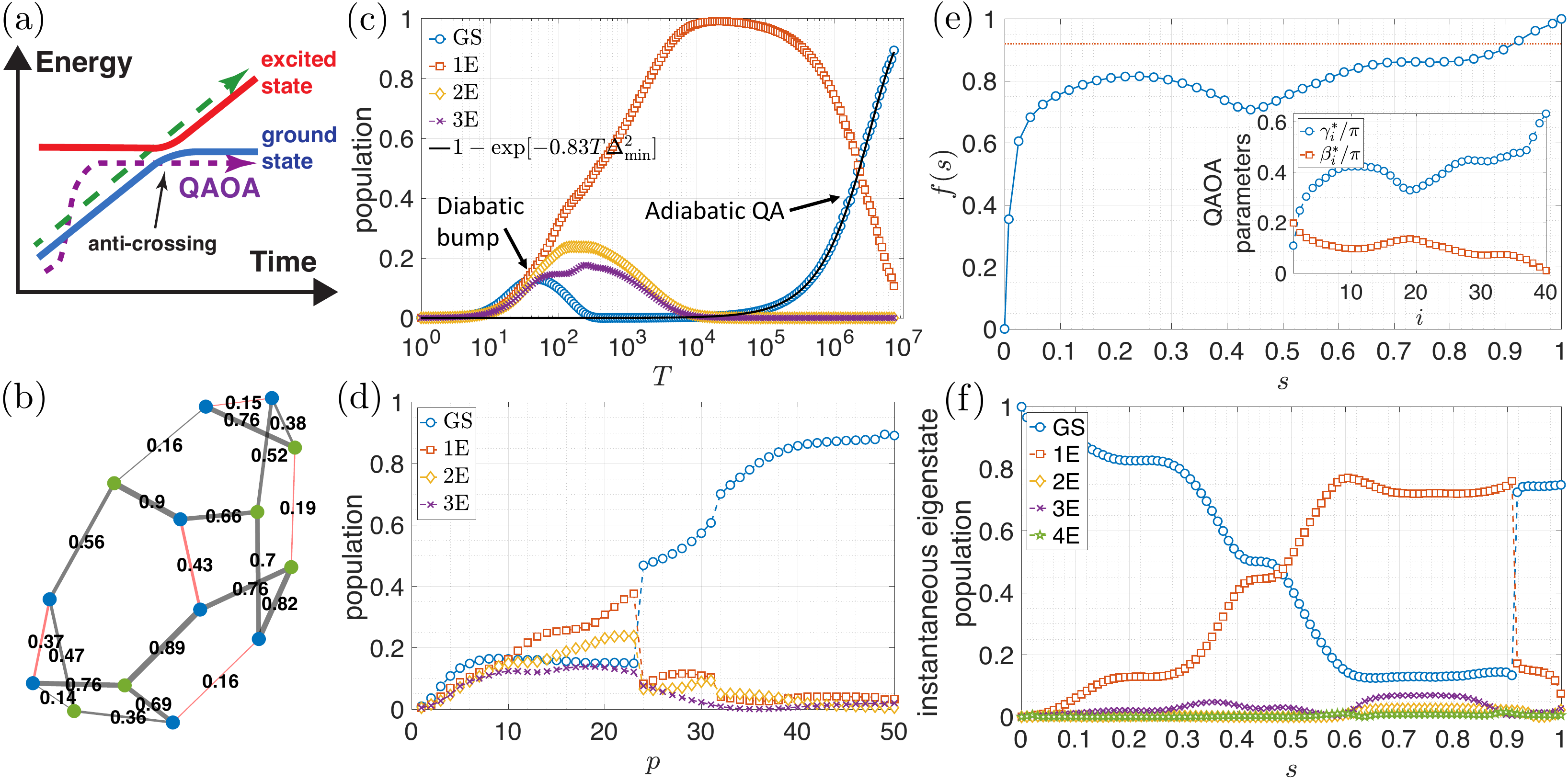} 
\caption{
(a) A schematic of how QAOA and the interpolated annealing path can overcome the small minimum gap limitations via diabatic transitions (purple line). Naive adiabatic quantum annealing path leads to excited states passing through the anti-crossing point (green line).
(b) An instance of weighted 3-regular graph that has a small minimum spectral gap along the quantum annealing path given by Eq.~\eqref{Eq:HamQA}. The optimal MaxCut configuration is shown with two vertex colors, and the black (red) lines are the cut (uncut) edges.
(c) Population in low excited states after the quantum annealing protocol with different total time $T$. The black solid line follows the Landau-Zener formula for the ground state population, $p_{\text{GS}} = 1- \exp\left( - c T \Delta_{\min}^{2}\right)$, where $c$ is a fitting parameter. 
(d) Population in low excited states using QAOA at different level $p$. The FOURIER heuristic strategy is used in the optimization.
(e) Interpreting QAOA parameters (at $p=40$) as a smooth quantum annealing path, via linear interpolation according to Eq.~\eqref{Eq:QAOAconversion}. The annealing time parameter $s = t/T_{p}$, where $T_{p} = \sum_{i=1}^{p} (|\gamma_{i}^{*}| + |\beta_{i}^{*}|)$. The red dotted line labels the location of anti-crossing where the gap is at its minimum, at which point $f(s) \approx 0.92$.  Inset shows the original QAOA optimal parameters $\gamma^{*}_{i}$ and $\beta^{*}_{i}$ for $p = 40$. 
(f) Instantaneous eigenstate population under the annealing path given in (e). Note that the instantaneous ground state and first excited state swap at the anti-crossing point.}
\label{Fig:QAOAvsQA}
\end{figure*}

To understand the behavior of QAOA, we focus on graph instances that are hard for adiabatic QA in this subsection.
In particular, we study a representative instance to explain how QAOA as well as diabatic QA can overcome the adiabatic limitations. As illustrated in Fig.~\ref{Fig:QAOAvsQA}(a), we demonstrate that QAOA can learn to utilize diabatic transitions at anti-crossings to circumvent difficulties caused by very small gaps in QA. 

Fig.~\ref{Fig:QAOAvsQA}(b) shows a representative graph with $N = 14$, whose minimum spectral gap $\Delta_{\min}<10^{-3}$. For this example hard instance, we first numerically simulate the quantum annealing process.
Fig.~\ref{Fig:QAOAvsQA}(c) shows populations in the ground state and low excited states at the end of the process for different annealing time $T$.
Since the minimum gap is very small, the adiabatic condition requires $T \gtrsim 1/ \Delta_{\min}^{2} \approx 10^{6}$.
The Landau-Zener formula for the ground state population $p_{\text{GS}} = 1- \exp\left( - c T \Delta_{\min}^{2}\right)$ fits well with our exact numerical simulation, where $c$ is a fitting parameter.
However, we can clearly observe a ``bump'' in the ground state population at annealing time $T \approx 40$.
At such a time scale, the dynamics is clearly non-adiabatic; we call this phenomenon a \emph{diabatic bump}.
This phenomenon has been observed earlier in Ref.~\cite{Crosson2014, Muthukrishnan2016Tunneling, Hormozi2017, Albash2018Adiabatic} for other optimization problems.

Subsequently, we simulate QAOA on this hard instance.
As mentioned earlier, although  QAOA optimizes energy instead of ground state overlap, substantial ground state population can still be obtained even for many hard graphs.
Using our FOURIER heuristic strategy, various low-energy state populations of the output state are shown for different levels $p$ in Fig.~\ref{Fig:QAOAvsQA}(d).
We observe that QAOA can achieve similar ground state population as the diabatic bump at small $p$, and then substantially enhance it after $p\gtrsim24$.

To better understand the mechanism of QAOA and make a meaningful comparison with QA, we can interpret the QAOA parameters as a smooth annealing path.
We again interpret the sum of the variational parameters to be the total annealing time, i.e., $T_{p} = \sum_{i=1}^{p} (|\gamma_{i}| + |\beta_{i}|)$.
A smooth annealing path can be constructed from QAOA optimal parameters as 
\begin{align}
& H_{\text{QAOA}}(t) = -[f(t) H_{C} + (1-f(t)) H_{B}], 
\label{Eq:QAOAconversion}
\\
& f \left( t_{i} = \sum_{j=1}^{i} (|\gamma_{j}^{*}| + |\beta_{j}^{*}|) - \dfrac{1}{2}(|\gamma_{i}^{*}| + |\beta_{i}^{*}|)  \right) = \dfrac{\gamma_{i}^{*}}{|\gamma_{i}^{*}| + |\beta_{i}^{*}|}, \nonumber
\end{align}
where $t_{i}$ is chosen to be at the mid-point of each time interval $(\gamma_{i}^{*} + \beta_{i}^{*})$.
With the boundary conditions $f(t=0) = 0$, $f(t=T_{p}) = 1$ and linear interpolation at other intermediate time $t$, we can convert QAOA parameters to a well-defined annealing path.
We apply this conversion to the QAOA optimal parameters at $p = 40$ [Fig.~\ref{Fig:QAOAvsQA}(e)].
With this annealing path, we can follow the instantaneous eigenstate population throughout the quantum annealing process [Fig.~\ref{Fig:QAOAvsQA}(f)].
In contrast to adiabatic QA, the state population leaks out of the ground state and accumulates in the first excited state before the anti-crossing point, where the gap is at its minimum.
Using a diabatic transition at the anti-crossing, the two states swap populations, and a large ground state population is obtained in the end.
We note that the final state population from the constructed annealing path differs slightly from those of QAOA, due to Trotterization and interpolation, but the underlying mechanism is the same, which is also responsible for the diabatic bump seen in Fig.~\ref{Fig:QAOAvsQA}(c).
In addition to the conversion used in Eq.~\eqref{Eq:QAOAconversion}, we have tested a few other prescriptions to construct an annealing path from QAOA parameters, and qualitative features do not seem to change.

Hence, our results indicate that QAOA is closely related to a cleverly optimized diabatic QA path that can overcome limitations set by the adiabatic theorem. Through optimization, QAOA can find a good annealing path and exploit diabatic transitions to enhance ground state population.
This explains the observation in Fig.~\ref{Fig:TTS}(a) that $\text{TTS}_{\text{QAOA}}^{\text{opt}}$ can be significantly shorter than the time required by the adiabatic algorithm.
On the other hand, as seen in Fig.~\ref{Fig:TTS}(b), $\text{TTS}_{\text{QAOA}}^{\text{opt}}$ is strongly correlated with $\text{TTS}_{\text{QA}}^{\text{opt}}$: QAOA automatically finds a good annealing path, which could be adiabatic or not, depending on what is the best route for the specific problem instance.   

The effective dynamics of QAOA for this specific instance, as we see in Fig.~\ref{Fig:QAOAvsQA}(f), can be understood mostly by an effective two-level system (see Appendix~\ref{sec:few-level} for more discussions).
In general, the energy spectrum can be more complex, and the dynamics may involve many excited states, which may not be explainable by the simple schematic in Fig.~\ref{Fig:QAOAvsQA}(a).
Nonetheless, QAOA can find a suitable path via our heuristic optimization strategies even in more complicated cases~\cite{Pichler2018Quantum}.


\vspace{-4pt}
\section{Considerations for experimental implementation}
\label{Sec:Experiment}

\vspace{-4pt}

In this section, we discuss some important considerations for experimental realization. The framework of QAOA is general and can be applied to various experimental platforms to solve combinatorial optimization problems. Here, we again focus on the MaxCut problem as a paradigmatic example, although it can also be applied to solve other interesting problems \cite{Pichler2018Quantum, Pichler2018Computational}.

\subsection{Finite measurement samples}

So far, we have focused on understanding the best theoretically possible performance of QAOA, and assumed perfect measurement precision of the objective function in our numerical simulations.
However, due to quantum fluctuations (i.e., projection noise) in actual experiments, the precision is finite since it is obtained via averaging over finitely many measurement outcomes that can only take on discrete values.
Hence, in practice, there is a trade-off between measurement cost and optimization quality: finding good optimum requires good precision at the cost of large number of measurements \cite{Guerreschi2017Practical}.
Additionally, large variance in the objective function value demands more measurements, but may help improve the chances of finding near-optimal MaxCut configurations.

\begin{figure}[t]
\includegraphics[width=\linewidth]{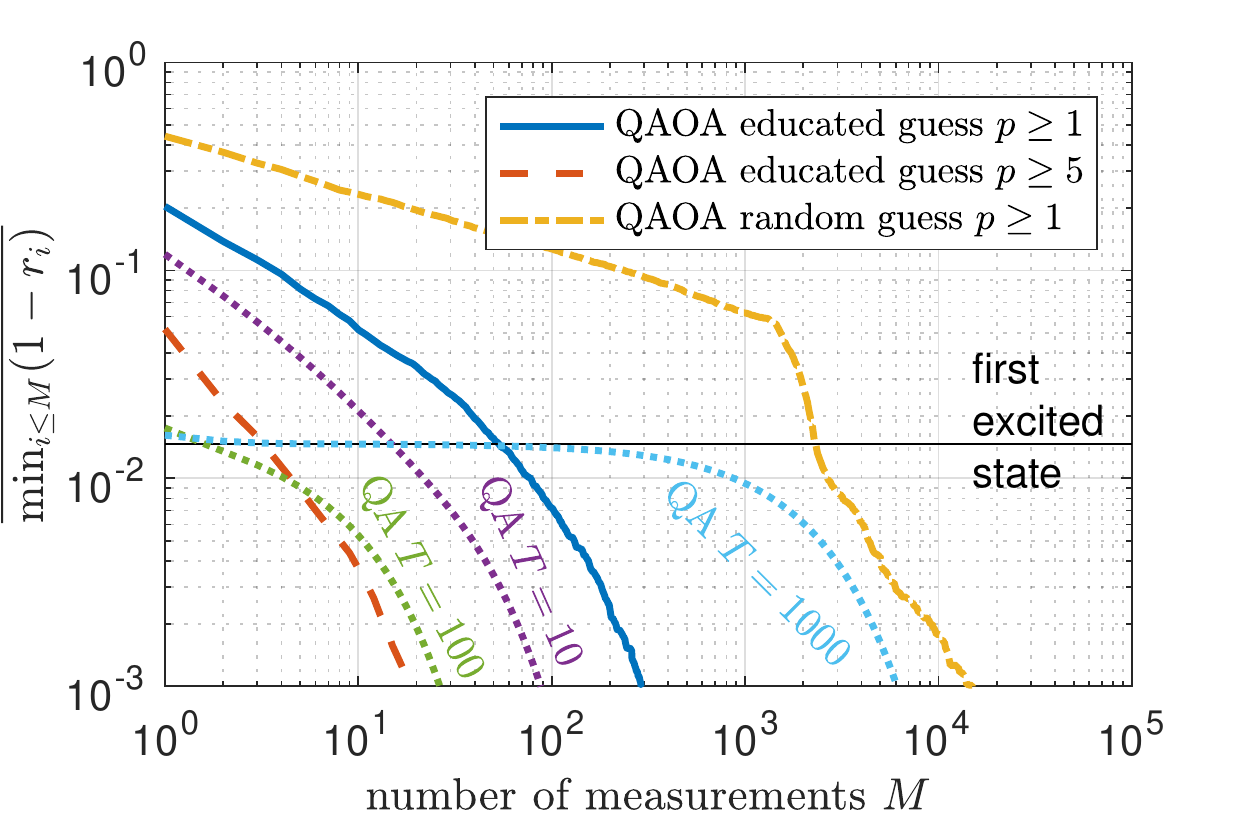}
\caption{\label{Fig:ShotNoise}
Full Monte-Carlo simulation of QAOA accounting for measurement projection noise, on the example 14-vertex instance studied in Fig.~\ref{Fig:QAOAvsQA}(b).
For various optimization strategies, we keep track of approximation ratio $r_i = |\text{Cut}_i|/|\text{MaxCut}|$ from the $i$-th measurement, and plot minimum fractional error $1-r_i$ found after $M$ measurements, averaged over many Monte-Carlo realizations.
The blue solid and red dashed lines correspond to QAOA optimized with the FOURIER strategy starting with an educated guess of $(\vgamma,\vbeta)$ at $p=1$ and $p=5$, respectively.
The orange dash-dot line corresponds to QAOA optimized starting with random guesses at $p=1$.
The three labelled dotted lines are results from QA with total time $T=10,10^2,10^3$.
}
\end{figure}

Here, we study the effect of measurement projection noise with a full Monte-Carlo simulation of QAOA on some example graphs, where objective function is evaluated by repeated projective measurements until its error is below a threshold.
More implementation details of this numerical simulation are discussed in Appendix~\ref{appx:shot-noise}.
In Fig.~\ref{Fig:ShotNoise}, we present the Monte-Carlo simulation result for the example instance studied earlier in Fig.~\ref{Fig:QAOAvsQA}(b).
Here, we simulate QAOA by starting at either level $p$\,=\,1 or $p$\,=\,5, and increasing to higher $p$ using our FOURIER heuristics. 
The initial parameters at $p$\,=\,1 and $p$\,=\,5, respectively, are chosen based on known optima found for smaller size instances.
We see that QAOA can find the MaxCut solution in $\sim$10--$10^2$ measurements, and starting our optimization at intermediate level ($p$\,=\,5) is better than starting at the lowest level ($p$\,=\,1).
In comparison, random choices of initial parameters starting at $p=1$ perform much worse, which fails to find the MaxCut solution until $10^3$--$10^4$ measurements have been made.
Moreover, when we compare QAOA to QA with various annealing time, it appears that the choice of annealing time $T=100$ can perform just as well as QAOA on this instance.
Nevertheless, running QAOA at level $p=5$ is still more advantageous than QA at $T=100$, when coherence time is limited.

We remark that our simulation is only limited to small-size instances, and the good performance of QAOA and QA we observe is complicated by the small but significant ground state population from generic annealing schedules.
Since it often takes $10^2$ measurements to obtain a sufficiently precise estimate of the objective function, a ground state probability of $\gtrsim10^{-2}$ would mean that one can find the ground state without much parameter optimization.
Nevertheless, as quantum computers with $10^2$\,$\sim$\,$10^3$ qubits begin to come online,
it will be interesting to see how QAOA performs on much larger instances where the ground state probability from generic QAOA/QA protocol is expected to be exponentially small, whereas the number of measurements necessary for optimization grows only polynomially with problem size.
The results here indicate that the parameter pattern and our heuristic strategies are practically useful guidelines in realistic implementation of QAOA.

\subsection{Implementation for large problem sizes}

As experiments begin to test solving the MaxCut problems with quantum machines~\cite{Otterbach2017Unsupervised, Qiang2018Photonics},
limited quantum coherence time and graph connectivity will be among the biggest challenges. In terms of coherence time, QAOA is highly advantageous: the hybrid nature of QAOA as well as its short- and intermediate-depth circuit parametrization makes it ideal for near-term quantum devices.
In addition, we have demonstrated that QAOA is not generally limited by the small spectral gaps, which raises hopes to (approximately) solve interesting problems within the coherence time. 

What would be the necessary problem size to explore a meaningful quantum advantage?
We note that one of the leading exact classical solver, the BiqMac solver~\cite{Rendl2008Solving}, is able to solve MaxCut exactly up to $N \lesssim 100$, but takes a long time (more than an hour) for larger problem sizes.
A fast heuristic algorithm, the breakout local search algorithm~\cite{Benlic2013Breakout}, can find the MaxCut solution with a high probability for problem size of a few hundreds, although the solution is not guaranteed.
Hence, a MaxCut problem with a few hundred vertices will be an interesting regime to benchmark quantum algorithms.
In terms of approximation, we have noted earlier that the polynomial-time Goemans-Willamson algorithm has an approximation ratio guarantee of $r^{*} \approx 0.878$.
It will be also interesting to find out if QAOA is able to achieve a better approximation ratio for some problem instances where the exact solution is not obtainable, in this regime of hundreds of vertices.

We now discuss a few considerations that put these large-size problems in the experimentally feasible regime on near-term quantum systems.

\begin{figure}[t]
\includegraphics[width=\columnwidth]{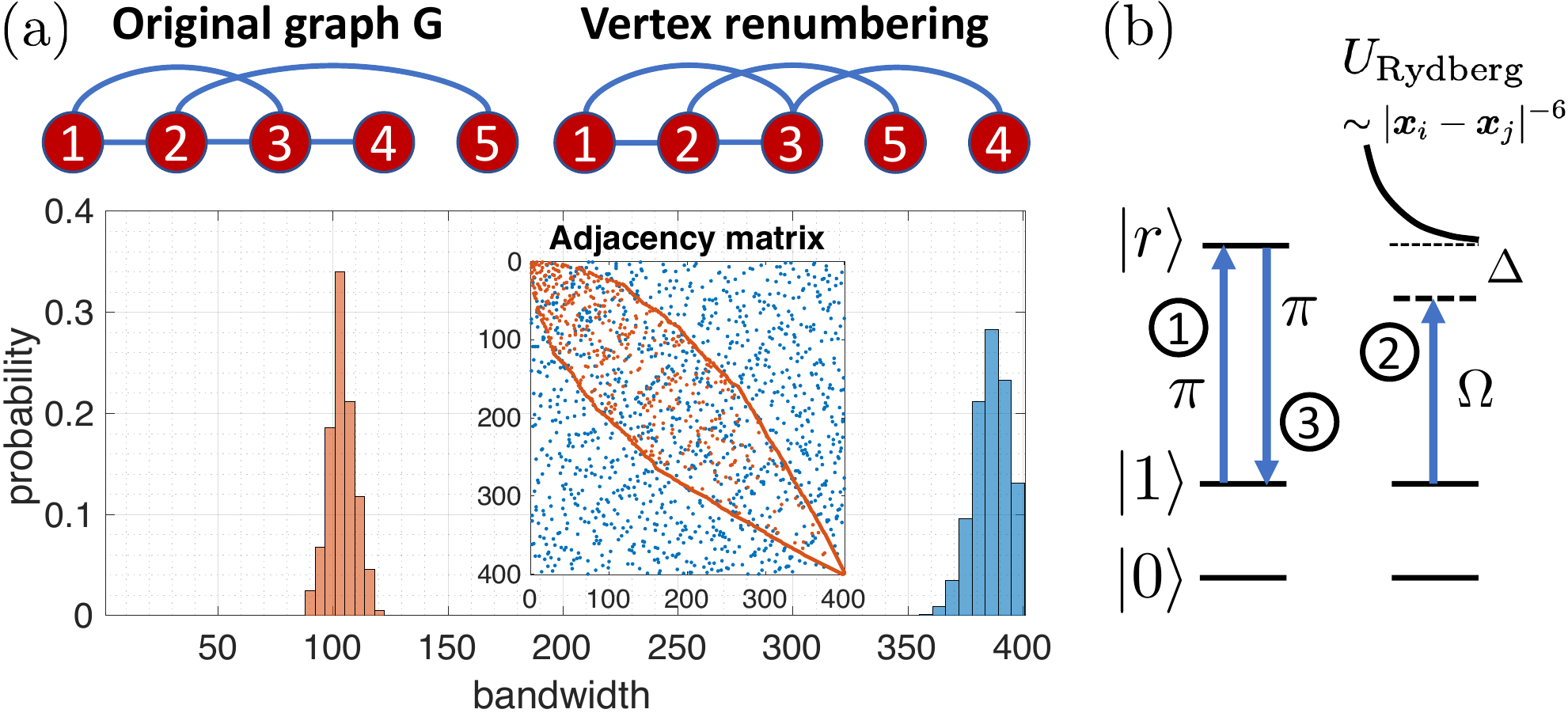} 
\caption{
(a) Vertex renumbering to reduce the graph bandwidth. Top panel: an example of vertex renumbering for a 5-vertex graph. Bottom panel: distribution of graph bandwidths for 1000 random 3-regular graphs with $N=400$ each. The blue (orange) bars are the bandwidth before (after) vertex renumbering. Inset shows the sparsity pattern of the adjacency matrix before and after vertex renumbering for one particular graph.
(b) A protocol to use Rydberg-blockade controlled gate to implement the interaction term $ \exp ( -i \gamma \sigma_{i}^{z} \sigma_{j}^{z})$. By choosing a proper gate time for the second step $\left( t = 2\pi/\sqrt{\Omega^{2}+\Delta^{2}} \right)$, one does not accumulate population on the Rydberg level $\ket{r}$. With tunable coupling strength $\Omega$ and detuning $\Delta$, one can control the interaction time $\gamma$.}
\label{Fig:Exp}
\end{figure}

\subsubsection{Reducing interaction range}

In a quantum experiment, each vertex can be represented by a qubit.
For a large problem size, a major challenge to encode general graphs will be the necessary range and versatility of the interaction patterns (between qubits).
The embedding of a random graph into a physical implementation with a 1D or 2D geometry may require very long-range interactions.
By relabelling the graph vertices, one can actually reduce the required range of interactions.
This can be formulated as the \emph{graph bandwidth problem}:
Given a graph $G = (V,E)$ with $N$ vertices, a vertex numbering is a bijective map from vertices to distinct integers, $f: V \to \{1,2,\cdots \!,N\}$. The bandwidth of a vertex numbering $f$ is $B_{f}(G) = \max\{|f(u) - f(v)|: (u,v) \in E(G) \}$, which can be understood as the length of the longest edge (in 1D).
The graph bandwidth problem is then to find the minimum bandwidth among all vertex numberings, i.e., $B(G) = \min_{f} B_{f}(G)$;
namely, it is to minimize the length of the longest edge by vertex renumbering.

In general, finding the minimum graph bandwidth is NP-hard, but good heuristic algorithms have been developed to reduce the graph bandwidth.
Fig.~\ref{Fig:Exp}(a) shows a simple example of bandwidth reduction. The top panel illustrates the vertex renumbering with a $5$-vertex graph. The bottom panel shows the histogram of graph bandwidths for 1000 random 3-regular graphs of $N = 400$ each.
Using the Cuthill-McKee algorithm~\cite{Cuthill1969Reducing}, the graph bandwidth can be reduced to around $ B \approx 100$.
While this still requires quite a long interaction range in 1D, the bandwidth problem can also be generalized to higher dimensions.
In 2D arrangements, we then expect the diameter of interaction to be $ B_{\text{2D}} \sim \sqrt{100}$ for $N=400$ 3-regular graphs~\cite{Lin2011Square}, which seems within reach for near-term quantum devices.
A detailed study of the 2D bandwidth problem is beyond the scope of this work. Alternatively, one can make use of a general construction to encode any long-range interactions to local fields~\cite{Lechner2015Quantum}, although it requires additional physical qubits and gauge constraints. Apart from implementing arbitrary graphs in full generality, one can also restrict to special graphs that exhibit some geometric structures.
For example, unit disk graphs are geometric graphs in the 2D plane, where vertices are connected by an edge only if they are within a unit distance.
These graphs can be more naturally encoded into 2D physical implementations, and the MaxCut problem is still NP-hard on unit disk graphs~\cite{Diaz2007MaxCut}.

\subsubsection{Example Implementation with Rydberg Atoms}

The above discussion has been experimental-platform independent, and is applicable to any state-of-the-art platforms, such as neutral Rydberg atoms~\cite{Saffman2010Quantum, Labuhn2016Tunable, Bernien2017Probing, Keesling2018Probing, Kumar2018Sorting}, trapped ions~\cite{Haffner2008Quantum, Debnath2016Demonstration, Zhang2017Observation, Kokail2018Self}, and superconducting qubits~\cite{Barends2016Digitized, Kandala2017Hardware, Otterbach2017Unsupervised, Neill2018Blueprint}.
As an example, we briefly describe a feasible implementation of QAOA with neutral atoms interacting via Rydberg excitations,
where high-fidelity entanglement has been recently demonstrated~\cite{Levine2018HighFidelity}.
We can use the hyperfine ground states in each atom to encode the qubit states $\ket{0}$ and $\ket{1}$, and the $\ket{1}$ state can be excited to the Rydberg level $\ket{r}$ to induce interactions.
The qubit rotating term, $\exp \left(-i \beta \sum_{j=1}^N \sigma_j^x \right)$, can be implemented straightforwardly by a global driving beam with tunable durations.
The interaction terms $ \sum_{\left<i,j \right>} \sigma_{i}^{z} \sigma_{j}^{z}$ can be implemented stroboscopically for general graphs;
this can be realized by a Rydberg-blockade controlled gate~\cite{Saffman2010Quantum}, as illustrated in Fig.~\ref{Fig:Exp}(b).
By controlling the coupling strength $\Omega$, detuning $\Delta$, and the gate time, together with single-qubit rotations, one can implement $ \exp ( -i \gamma \sigma_{i}^{z} \sigma_{j}^{z})$, which can then be repeated for each interacting pair. 
An additional major advantage of the Rydberg-blockade mechanism is its ability to perform multi-qubit collective gates in parallel~\cite{Saffman2010Quantum, Isenhower2011Multibit}. This can reduce the number of two-qubit operation steps from the number of edges to the number of vertices, $N$, which means a factor of $N$ reduction for dense graphs with $\sim$$N^{2}$ edges. While the falloff of Rydberg interactions may limit the distance two qubits can interact, MaxCut problems of interesting sizes can still be implemented by vertex renumbering or focusing on unit disk graphs, as discussed above.

For generic problems of 400-vertex 3-regular graphs, we expect the necessary interaction range to be $5$ atoms in 2D;
assuming minimum inter-atom separation of 2~$\mu$m, this means an interaction radius of 10~$\mu$m, which is realizable with high Rydberg levels~\cite{Saffman2010Quantum}.
Given realistic estimates of coupling strength $\Omega\sim 2\pi \times 10$-100~MHz and single-qubit coherence time $\tau\sim$ 200~$\mu$s (limited by Rydberg level lifetime), we expect with high-fidelity control, the error per two-qubit gate can be made roughly $(\Omega\tau)^{-1}$\,$\sim$\,$10^{-3}$-$10^{-4}$.
For 400-vertex 3-regular graphs, we can implement QAOA of level $p\simeq \Omega\tau/N \sim 25$ with a 2D array of neutral atoms.
We note that these are conservative estimates since we do not consider advanced control techniques such as pulse-shaping, and require less than one error in the entire quantum circuit;
it is possible that QAOA is not sensitive to some of the imperfections.
Hence, we envision that soon, QAOA can be benchmarked against classical~\cite{GW, Rendl2008Solving, Benlic2013Breakout} and semiclassical~\cite{Inagaki2016Coherent, McMahon2016Fully, Hamerly2018Experimental} solvers on large problem sizes with near-term quantum devices.


\section{Conclusion and Outlook}

In summary, we have studied the performance, non-adiabatic mechanism, and practical implementation of QAOA applied to MaxCut problems. 
Our results provide important insights into the performance of QAOA, and suggest promising strategies for its practical realization on near term quantum devices.
Based on the observed patterns in the QAOA optimal parameters, 
%
we developed heuristic strategies for initializing optimization so as to efficiently find quasi-optimal parameters in $O(\poly(p))$ time.
In contrast, optimization with the standard random initialization strategy that explores the entire parameter space needs $2^{O(p)}$ runs to obtain an equally good solution.
We benchmarked, using these heuristic optimization strategies, the performance of QAOA up to $p \leq 50$.
On average, the performance characterized by the approximation ratio was observed to improve exponentially (or stretched-exponentially) for random unweighted (or weighted) graphs.
Focusing on hard graph instances where adiabatic QA fails due to extremely small spectral gaps, we found that QAOA could learn via optimization a diabatic path to achieve significantly higher success probabilities to find the MaxCut solution.
A metric taking into account of both the success probability and the run time indicates QAOA may not be limited by small spectral gaps. 
Finally, we provided a detailed resource analysis for experimental implementation of QAOA on MaxCut and proposed a neutral-atom realization where problem sizes of a few hundred atoms are feasible in the near future. 

As we have benchmarked via random MaxCut instances, our simple heuristic optimization strategies work very well.
Nevertheless, more sophisticated methods could be developed to improve the performance and robustness.
One possibility would be using machine learning techniques to automatically learn and make use of the optimal parameter patterns to develop even more efficient parametrization and strategies (see e.g.,~\cite{Bukov2018Reinforcement}).
Another important but unsettled problem is assessing a reliable system-size scaling for QAOA.
On average, we observed (stretched) exponential improvement with level $p$, up to $ N = 22$.
It remains open whether the same scaling will persist for larger system size.
For the hard instances we generated that have exceedingly small spectral gaps, QAOA is able to overcome the adiabatic limitations in all cases examined;
it remains to be seen how this behavior could extrapolate to larger problems.
An experimental implementation with near-term quantum devices will be able to push the limit of our understanding and serve as a litmus test for genuine quantum speedup in solving practical problems. 

Besides MaxCut, another interesting optimization problem is the Maximum Independent Set (MIS) problem, which is also NP-hard and has many real-world applications~\cite{Vahdatpour:2008hb,Agarwal:1998fn}.
In a separate work~\cite{Pichler2018Quantum}, we show that the MIS problem can be naturally encoded into the ground state of neutral atoms interacting via Rydberg excitations, with minimal overhead on the hardware.
The methodology we have developed here for QAOA on MaxCut can be directly adapted to the MIS problem, where we observe similar parameter patterns and non-adiabatic mechanisms of QAOA; this is discussed in more detail in Appendix~\ref{sec:MIS}.
With such rapid development in near-term quantum computer, we will soon be able to witness experimental tests of the capability of quantum algorithms to solve practically interesting problems.

\begin{acknowledgments}

We thank Edward Farhi, Aram Harrow, Peter Zoller, Ignacio Cirac, Jeffrey Goldstone, Sam Gutmann, Wen Wei Ho, and Dries Sels for fruitful discussions. S.-T.W. in addition would like to thank Daniel Lidar, Matthias Troyer, and Johannes Otterbach for insightful discussions and suggestions during the Aspen Winter Conference on Advances in Quantum Algorithms and Computation. This work was supported through the National Science Foundation (NSF), the Center for Ultracold Atoms, the Air Force Office of Scientific Research via the MURI, the Vannevar Bush Faculty Fellowship, DOE, and Google Research Award. S.C. acknowledges support from the Miller Institute for Basic Research in Science. H.P. is supported by the NSF through a grant for the Institute for Theoretical Atomic, Molecular, and Optical Physics at Harvard University and the Smithsonian Astrophysical Observatory. Some of the computations in this paper were performed on the Odyssey cluster supported by the FAS Division of Science, Research Computing Group at Harvard University. \\

\noindent\emph{Note added.}---After the completion of this work, we became aware of a related work appearing in Ref.~\cite{Crooks2018Performance}.
In the preprint, they trained the QAOA variational parameters on a batch of graph instances and compared QAOA's performance with the classical Goemans-Williamson algorithm on small-size MaxCut problems.
Similar parameter shapes were found, but Ref.~\cite{Crooks2018Performance} did not make use of any observed patterns to design optimization strategies. We, in addition, discovered non-adiabatic mechanisms of QAOA, which is not studied in Ref.~\cite{Crooks2018Performance}.

\end{acknowledgments}

\bibliography{QAOArefs}


\appendix

\begin{figure}[b]
\includegraphics[width=\linewidth]{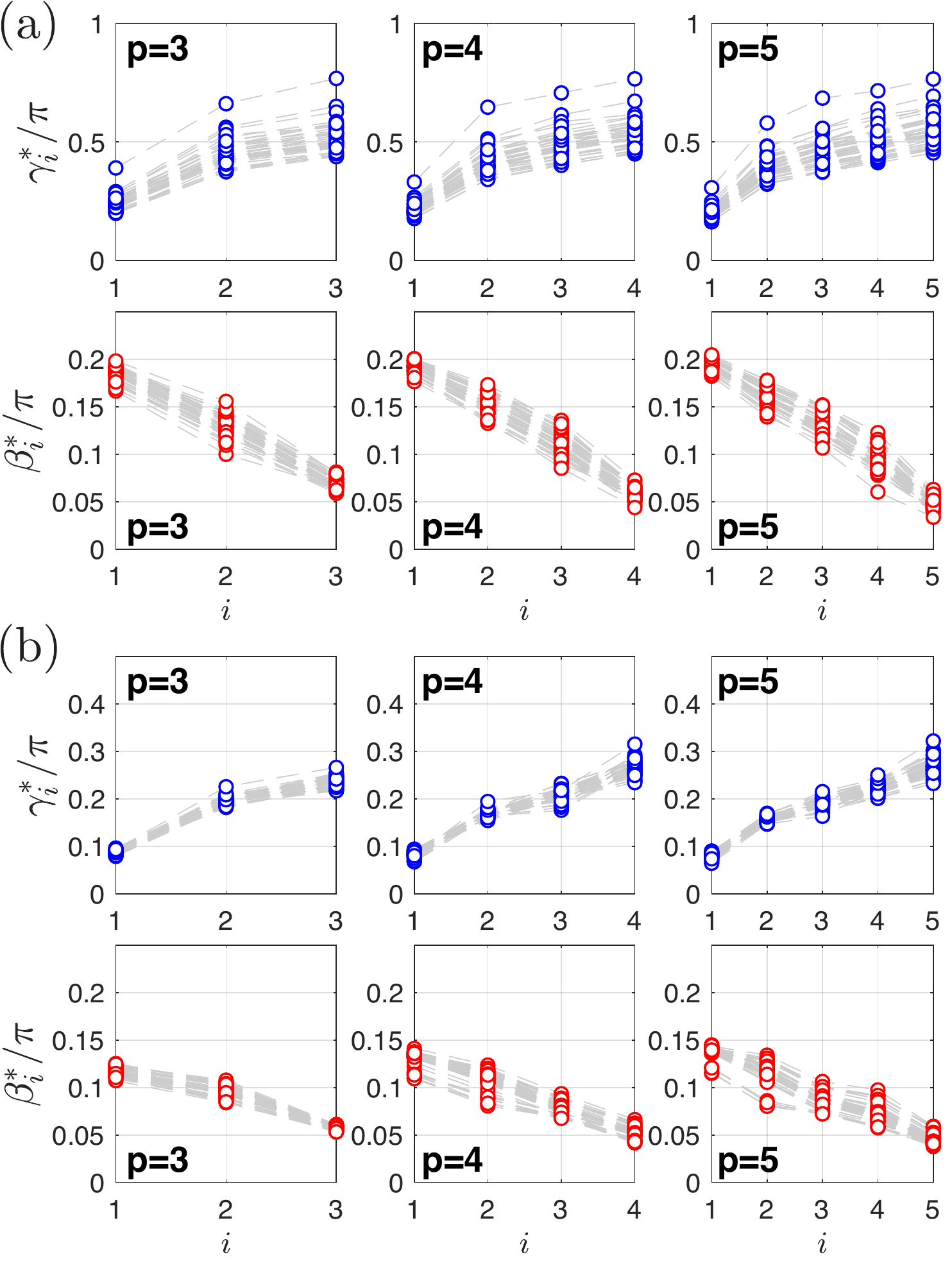} 
\caption{Optimal QAOA parameters $(\vgamma^{*},\vbeta^{*})$ for 40 instances of 16-vertex (a) weighted 3-regular (w3R) graphs and (b) weighted complete graphs, for $3\le p \le 5$. Each dashed line connects parameters for one particular graph instance. For each instance and each $p$, the BFGS routine is used to optimize from $10^4$ random initial points, and the best parameters are kept as $(\vgamma^*, \vbeta^*)$.}
\label{fig:param-pattern-weighted}
\end{figure}

\section{Optimal parameter pattern for weighed graphs\label{appx:param-pattern}}

Here, we illustrate the pattern of optimal QAOA parameters for both weighted 3-regular (w3R) graphs and weighted complete graphs.
The weight of each edge is randomly drawn from uniform distribution on the interval $[0,1]$.
In Fig.~\ref{fig:param-pattern-weighted}, we illustrate the pattern by simultaneously plotting the optimal parameters for 40 instances.
In both cases, we see a pattern analogous to what was found for unweighted 3-regular (u3R) graphs in Fig.~\ref{fig:param-pattern}(b), where $\gamma_i^*$ tend to increase smoothly and $\beta_i^*$ tend to decrease smoothly with $i=1,2,\ldots, p$.
The optimal parameter of graphs from the same class also appears to cluster together in the same range.

We observe that the spread of $\vgamma^{*}$ for w3R graphs in Fig.~\ref{fig:param-pattern-weighted}(a) is wider than that for u3R graphs in Fig.~\ref{fig:param-pattern}(b). 
This is because the random weights effectively increase the number of subgraph types.
Moreover, the larger value for $\vgamma^{*}$ for w3R compared to u3R graphs can be understood via the effective mean-field strength that each qubit experiences.

For complete graphs in Fig.~\ref{fig:param-pattern-weighted}(b), we observe that $\vgamma^*$ for different weighted graphs have a narrower spread as well as smaller value compared to both u3R and w3R graphs. This is because there is only one type of subgraph that every edge sees. The non-zero spread is attributed to the fact that there are random weights on the edges. We also expect this spread to vanish as problem size $N$ increases, when for a typical instance, the distribution of weights on the $N-1$ edges incident to each qubits converges.
The smaller value of $\vgamma^*$ can also be understood via the effective mean-field strength picture, as each qubit interacts with all $N-1$ other qubits.

\section{Details of heuristic optimization strategies\label{appx:heuristics}}

In the main text, we have proposed two classes of heuristic strategies called INTERP and FOURIER for generating initial points in optimizing QAOA parameters.
Both \mbox{INTERP} and FOURIER strategies work well for all the instances we have examined.
We have chosen to use FOURIER for the results in the main text due to the slight edge in its performance in finding better optima when random perturbations are introduced.
We now explain how these strategies work in details, and compare their performances.

\subsection{Interpolation-based strategy\label{appx:INTERP}}

In the optimization strategy that we call INTERP, we use linear interpolation to produce a good initial point for optimizing QAOA as one iteratively increases the level $p$.
This is based on the observation that the shape of parameters $(\vgamma^*_{(p+1)},\vbeta^*_{(p+1)})$ closely resembles that of $(\vgamma^*_{(p)},\vbeta^*_{(p)})$.

The strategy works as follows: For a given instance, we iteratively optimize QAOA starting from $p=1$, and increment $p$ after obtaining a local optimum $(\vgamma_{(p)}^L, \vbeta_{(p)}^L)$. In order to optimize parameters for level $p+1$, we take the optimized parameters from level $p$ and produce initial points $(\vgamma^{0}_{(p+1)}, \vbeta^{0}_{(p+1)})$ according to:
\begin{align}
\left[\vgamma^{0}_{(p+1)}\right]_i & = \tfrac{i-1}{p} \left[\vgamma^{L}_{(p)}\right]_{i-1}+ \tfrac{p-i+1}{p}\left[\vgamma^{L}_{(p)}\right]_i
\end{align}
for $i=1,2,\ldots, p+1$.
Here, we denote $[\vgamma]_i\equiv \gamma_i$ as the $i$-th element of the parameter vector $\vgamma$, and $[\vgamma_{(p)}^{L}]_0\equiv [\vgamma_{(p)}^{L}]_{p+1} \equiv 0$. 
The expression for $\vbeta^{0}_{(p+1)}$ is the same as above after replacing $\gamma \rightarrow \beta$.
Starting from $(\vgamma^{0}_{(p+1)}, \vbeta^{0}_{(p+1)})$, we then optimize (e.g., using the BFGS routine) to obtain a local optimum $(\vgamma_{(p+1)}^L, \vbeta_{(p+1)}^L)$ for the $(p+1)$-level QAOA.
Finally, we increment $p$ by one and repeat the same process until a target level is reached.

\subsection{Details of FOURIER[$q,R$] strategy\label{appx:FOURIER}}

We now provide the details of our second heuristic strategy for optimizing QAOA that we call FOURIER.
As described in the Sec.~\ref{Sec:heuristic} main text, here we use a new representation of the $p$-level QAOA parameters as $(\vu,\vv)\in\mathds{R}^{2q}$ where
\begin{equation} \label{eq:uv-rep}
\begin{split}
\gamma_i &= \sum_{k=1}^q u_k \sin\left[\left(k-\frac12 \right)\left(i-\frac12 \right)\frac{\pi}{p}\right], \\
\beta_i  &= \sum_{k=1}^q v_k \cos\left[\left(k-\frac12\right)\left(i-\frac12 \right)\frac{\pi}{p}\right].\
\end{split}
\end{equation}
Roughly speaking, the FOURIER strategy works by starting with level $p=1$, optimize, and then reuse the optimum at level $p$ in $(\vu,\vv)$-representation to generate a good initial point for level $p+1$. 

In fact, we propose several variants of the FOURIER strategy for optimizing $p$-level QAOA: they are called FOURIER[$q,R$], characterized by two integer parameters $q$ and $R$.
The first integer $q$ labels the maximum frequency component we allow in the amplitude parameters $(\vu, \vv)$.
When we set $q = p$, the full power of $p$-level QAOA can be utilized,
in which case we simply denote the strategy as FOURIER[$\infty,R$], since $q$ grows unbounded with $p$.
Nevertheless, the smoothness of the optimal parameters $(\vgamma, \vbeta)$ implies that only the low-frequency components are important. Thus, we can also consider the case where $q$ is a fixed constant independent of $p$, so the number of parameters is bounded even as the QAOA circuit depth increases.
The second integer $R$ is the number of random perturbations we add to the parameters so that we can sometimes escape a local optimum towards a better one.
For the results shown in the main text, we use the FOURIER[$\infty,10$] strategy, meaning we set $q = p$ and $R=10$, unless otherwise stated.

In the basic FOURIER[$\infty,0$] variant of this strategy, we generate a good initial point for level $p+1$ by adding a higher frequency component, initialized at zero amplitude, to the optimum at level $p$.
More concretely, we take parameters from a local optimum $(\vu^L_{(p)}, \vv^L_{(p)})\in\mathds{R}^{2p}$ at level $p$, and generate a initial point $(\vu^0_{(p+1)}, \vv^0_{(p+1)})\in \mathds{R}^{2(p+1)}$ according to
\begin{align}
\vu^0_{(p+1)} = (\vu^L_{(p)}, 0), \quad
\vv^0_{(p+1)} = (\vv^L_{(p)}, 0).
\label{eq:FOURIER-R0}
\end{align}
Using $(\vu^0_{(p+1)}, \vv^0_{(p+1)})$ as a initial point, we perform BFGS optimization routine to obtain a local optimum $(\vu^L_{(p+1)}, \vv^L_{(p+1)})$ for the level $p+1$.

\begin{figure}[b]
\includegraphics[width=0.5\textwidth]{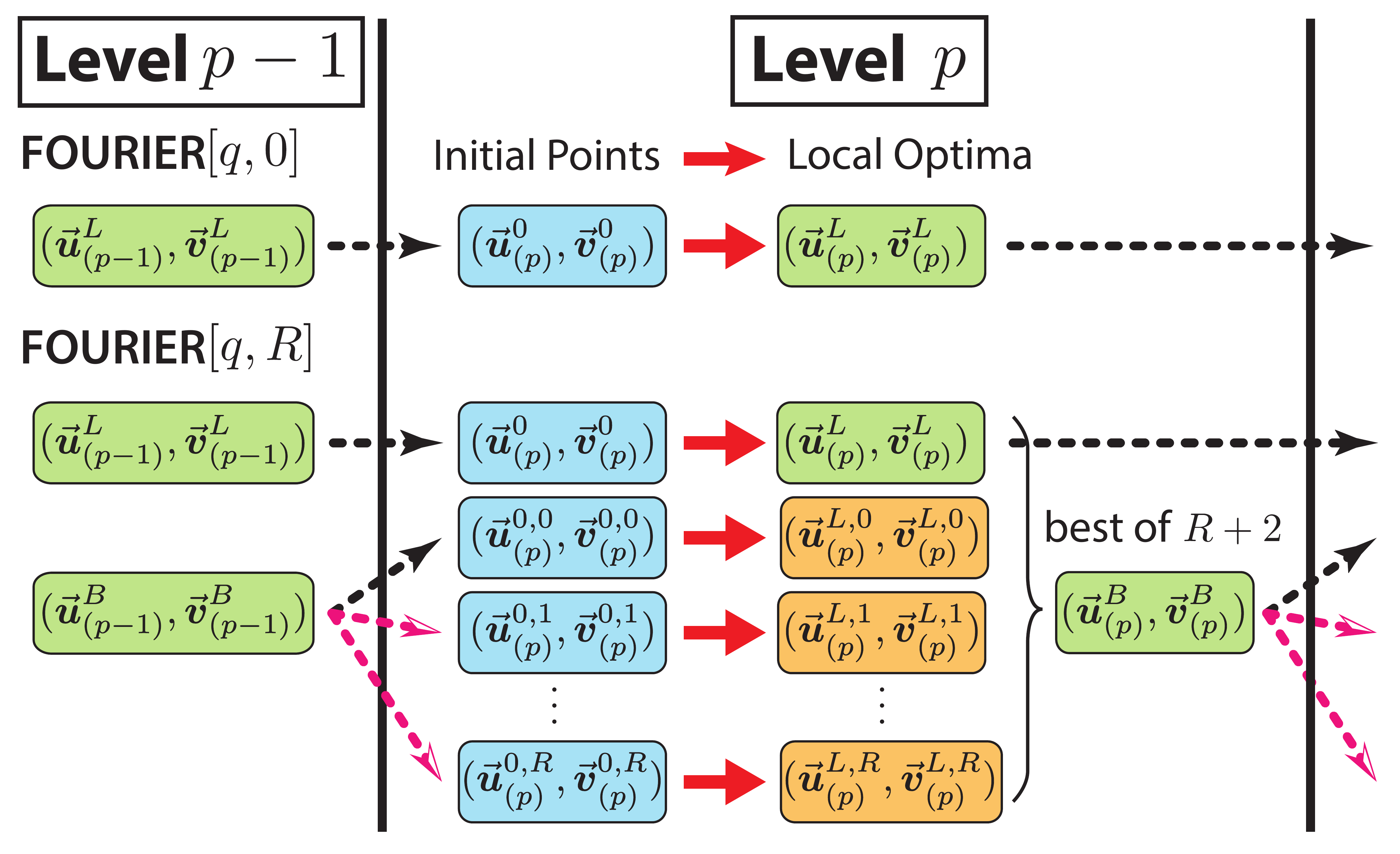}
\caption{\label{fig:heuristic-diagram}
Schematics of the two variants of FOURIER[$q,R$] heuristic strategy for optimizing QAOA, when $R=0$ and $R>0$.
Optimized parameters (green) at level $p-1$ are used to generate good initial points (blue) for optimizing at level $p$;
the same process is repeated to optimize for $p+1$.
When generating initial points, black dashed arrows indicate appending a higher frequency component [Eq.~\eqref{eq:FOURIER-R0}], and pink dashed arrows correspond to adding random perturbations [Eq.~\eqref{eq:FOURIER-RP}].
In the $R>0$ variant, two local optima (green) are kept in parallel at each level $p$ for generating good initial points: $(\vu_{(p)}^L, \vv_{(p)}^L)$ is the same optimum obtained from FOURIER[$q,0$] strategy, and 
$(\vu_{(p)}^B, \vv_{(p)}^B)$ is the best of $R+2$ local optima, $R$ of which are obtained from perturbed initial points.
We find that keeping the $(\vu_{(p)}^L, \vv_{(p)}^L)$ optimum improves the stability of the heuristic, as random perturbations can sometimes lead to erratic and non-smooth optimal parameters.
}
\end{figure}

We also consider an improved variant of the strategy, FOURIER[$\infty,R>0$], which is sketched alongside the $R=0$ variant in Fig.~\ref{fig:heuristic-diagram}.
This is motivated by our observation that the basic FOURIER[$\infty,0$] strategy can sometimes get stuck at a sub-optimal local optimum, whereas perturbing its initial point can improve the performance of QAOA.  
Here, in addition to optimizing according to the basic strategy, we optimize $(p+1)$-level QAOA from $R+1$ extra initial points, $R$ of which are generated by adding random perturbations to the best of all local optima $(\vu^B_{(p)},\vv^B_{(p)})$ found at level $p$.
Specifically, for each instance at $(p+1)$-level QAOA, and for $r=0,1,\ldots, R$, we optimize starting from
\begin{align}
\begin{split}
\vu^{0,r}_{(p+1)} &=
\begin{cases}
(\vu^B_{(p)}, 0),& r = 0\\
(\vu^B_{(p)} + \alpha \vu^{P,r}_{(p)}, 0),& 1\le r \le R
\end{cases}
\\
\vv^{0,r}_{(p+1)} &=
\begin{cases}
(\vv^B_{(p)}, 0),& r = 0\\
(\vv^B_{(p)} + \alpha \vv^{P,r}_{(p)}, 0), & 1\le r \le R
\end{cases}
\end{split}
\label{eq:FOURIER-RP}
\end{align}
where $\vu^{P,r}_{(p)}$ and $\vv^{P,r}_{(p)}$ contain random numbers drawn from normal distributions with mean 0 and variance given by $\vu^B_{(p)}$ and $\vv^B_{(p)}$:
\begin{align}
\begin{split}
\left[\vu^{P,r}_{(p)}\right]_k \sim \text{Normal}\left(0, \left[\vu^B_{(p)}\right]_k^2\right), \\
\left[\vv^{P,r}_{(p)}\right]_k \sim \text{Normal}\left(0, \left[\vv^B_{(p)}\right]_k^2\right).
\end{split}
\end{align}
There is a free parameter $\alpha$ corresponding to the strength of the perturbation.
Based on our experience from trial and error, setting $\alpha=0.6$ has consistently yielded good results. This choice of $\alpha$ is assumed for the results in this paper.

We remark that while the INTERP strategy can also get stuck in a local optimum, we find that adding perturbations to INTERP does not work as well as to FOURIER.
We attribute this to the fact that the optimal parameters are smooth, and adding perturbations in the $(\vu,\vv)$-space modify $(\vgamma,\vbeta)$ in a correlated way, which could enable the optimization to escape local optima more easily.
Hence, we consider here FOURIER with perturbations, but not INTERP.

Finally, we also consider variants of the FOURIER strategy where the number of frequency components $q$ is fixed.
These variants are the same as aforementioned strategies where $q=p$, except we truncated each of the $\vu$ and $\vv$ parameters to keep only the first $q$ components.
For example, when optimizing QAOA at level $p \ge q$ with the FOURIER[$q,0$] strategy, we stop adding higher frequency components and use the initial point $\vu^0_{(p+1)} = \vu^L_{(p)}\in \mathds{R}^{q}$.

\subsection{Comparison between heuristics}

\begin{figure}
\includegraphics[width=\linewidth]{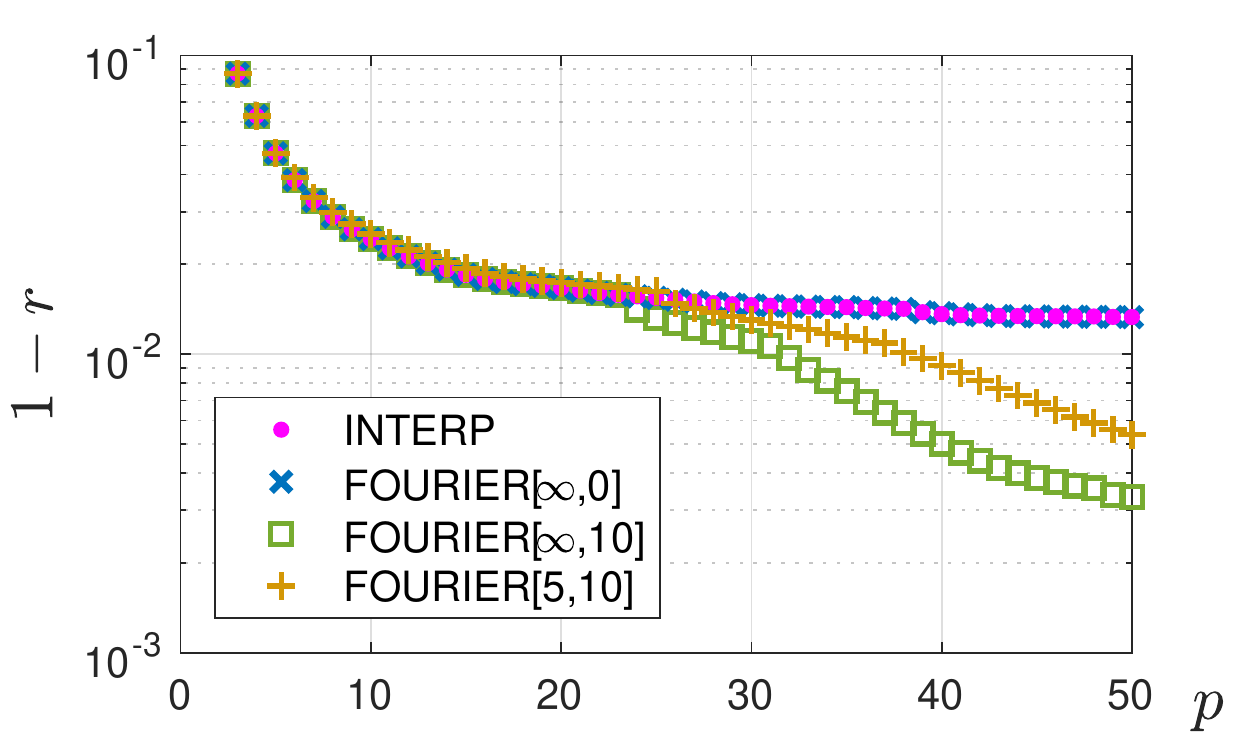}
\caption{\label{fig:comp-heu}The fractional error achieved by optimizing using our various heuristics on an example instance of 14-vertex w3R graph. This is the same instance shown in Fig.~\ref{Fig:QAOAvsQA}(b).}

\end{figure}

We now examine the difference in the performance among the various heuristic strategies we proposed.
For our example instance in Fig.~\ref{fig:comp-heu}, we note that \mbox{INTERP} and FOURIER[$\infty,0$] strategies have essentially identical performance (except for small variations barely visible at large $p$).
This is expected since both strategies generate initial points for optimizing level $p+1$ based on smooth deformation of the optimum at level $p$.
In any case, the FOURIER[$\infty,10$] strategy outperforms INTERP and FOURIER[$\infty,0$] beginning at level $p\gtrsim20$.
Interestingly, even when restricting the number of QAOA parameters to $2q=10$, the FOURIER[5,10] strategy closely matches the performance of other heuristics at low $p$ and beats the $R=0$ heuristic at large $p$.
This suggests that the optimal QAOA parameters may in fact live in a small-dimensional manifold.
Therefore, optimization for intermediate $p$-level QAOA can be dramatically simplified by using our new parameterization $(\vu,\vv$).

\section{Time-to-solution (TTS) for example graph instance}
\label{sec:ExampleGraph}

\begin{figure}[t]
\includegraphics[width=\linewidth]{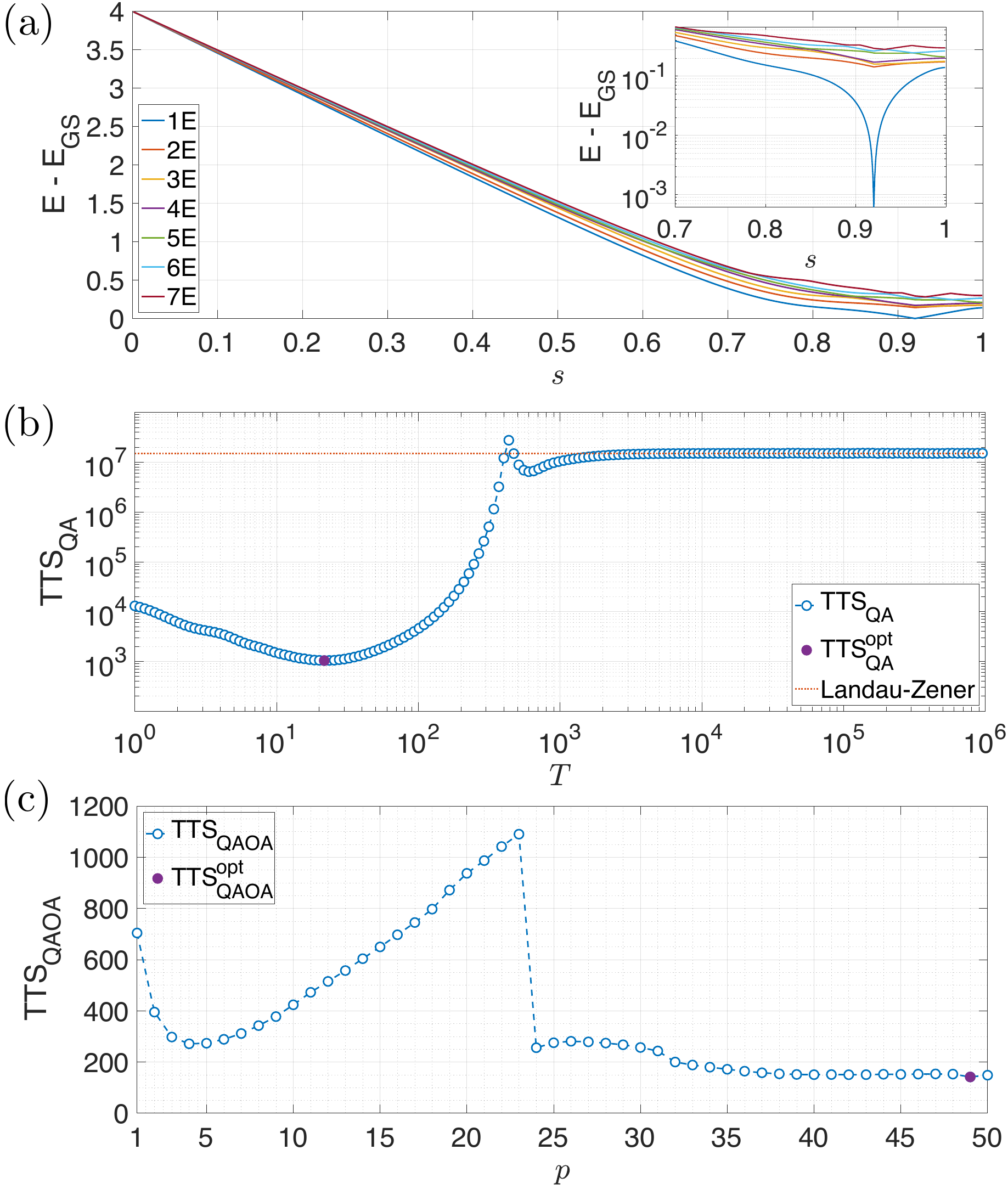} 
\caption{(a) Energy spectrum of low excited states (measured from the ground state energy $E_{\text{GS}}$) along the annealing path for the graph instance given in Fig.~\ref{Fig:QAOAvsQA}(b). Only states that can couple to the ground state are shown, i.e., states that are invariant under the parity operator $\mathcal{P} = \prod_{i=1}^{N} \sigma_{i}^{x}$. The inset shows the energy gap from the ground state in logarithmic scale.
(b) Time-to-solution for the linear-ramp quantum annealing protocol, $\text{TTS}_{\text{QA}}$, for the same graph instance. TTS in the long time limit follows a line predicted by the Landau-Zener formula, which is independent of the annealing time $T$.
(c) TTS for QAOA at each iteration depth $p$.}
\label{Fig:TTS-Example}
\end{figure}

In Sec.~\ref{subSec:mechanism}, we focused on a representative graph instance where the adiabatic minimum gap is small [Fig.~\ref{Fig:QAOAvsQA}(b)]. The low energy spectrum for the graph along the QA path can be seen in Fig.~\ref{Fig:TTS-Example}(a).
We remark that only eigenstates that are invariant under the parity operator $\mathcal{P} = \prod_{i=1}^{N} \sigma_{i}^{x}$ are shown, since the Hamiltonian $H_{\text{QA}}(s)$ commutes with $\mathcal{P}$ and the initial state is $\mathcal{P}$-invariant: $\mathcal{P}\ket{\psi(0)}=\ket{\psi(0)}$.

Fig.~\ref{Fig:TTS-Example}(b) and Fig.~\ref{Fig:TTS-Example}(c) illustrate $\text{TTS}_{\text{QA}}$ and $\text{TTS}_{\text{QAOA}}$ for the same graph.
For QA, one can see that non-adiabatic evolution with $T \approx 20$ yields orders of magnitude shorter TTS than the adiabatic evolution.
We also see that TTS in the adiabatic limit is independent of the annealing time $T$, following the Landau-Zener formula $p_{\text{GS}} = 1- \exp\left( - c T \Delta_{\min}^{2}\right)$.
For QAOA, we use our FOURIER[$\infty,10$] heuristic strategy to perform the numerical simulation up to $p_{\max} = 50$,
and compute $\text{TTS}_{\text{QAOA}}(p)$ and $\text{TTS}_{\text{QAOA}}^{\text{opt}}$ (up to $p\le p_{\max}$).
For this particular graph, we note that although $\text{TTS}_{\text{QAOA}}^{\text{opt}}$ occurs at $p = 49$, in practice it may be better to run QAOA at $p = 4$ or $p=5$ due to optimization overhead and error accumulation at deeper circuit depths.
The apparent discontinuous jump in $\text{TTS}_{\text{QAOA}}$ is due to the corresponding jump in $p_{\text{GS}}(p)$, which can be explained by two reasons:
first, our heuristic strategy is not guaranteed to find the global optimum, and random perturbations may help the algorithm escape a local optimum, resulting in a jump in ground state population;
second, even when the global optimum is found for all level $p$, there can still be discontinuities in $p_{\text{GS}}$, since the objective function of QAOA is energy instead of ground state population.

\section{Effective few-level understanding of the diabatic bump\label{sec:few-level}}

\begin{figure}[t]
\includegraphics[width=\columnwidth]{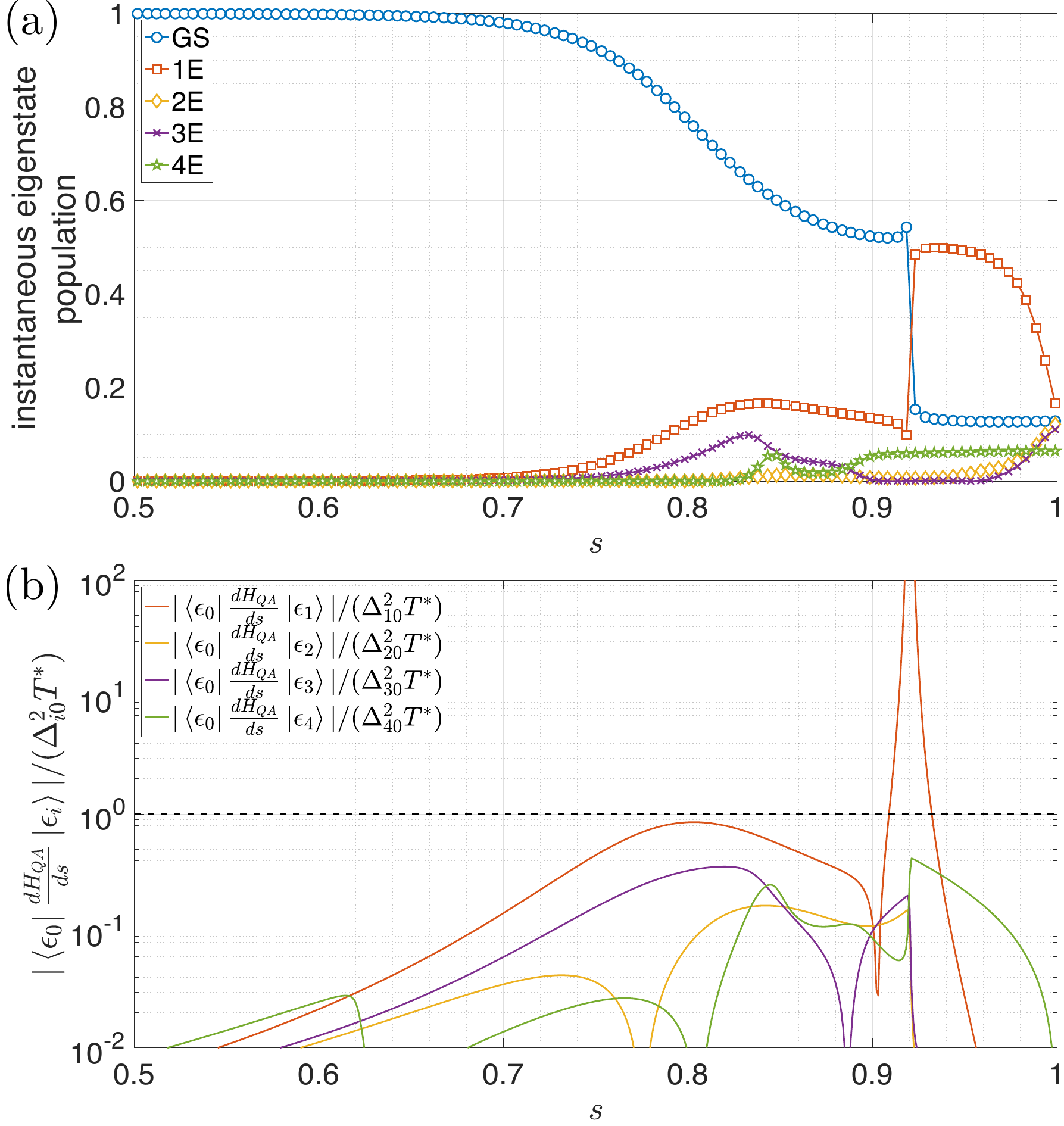} 
\caption{(a) Instantaneous eigenstate populations along the linear-ramp quantum annealing path given by Eq.~\eqref{Eq:HamQA} for the example graph in Fig.~\ref{Fig:QAOAvsQA}(b). The annealing time is chosen to be $T = T^{*} = 40$, which corresponds to the time where the diabatic bump occurs in Fig.~\ref{Fig:QAOAvsQA}(c). (b) Coupling between the instantaneous ground states and the first few excited states. The plotted quantities measure the degree of adiabaticity [as explained in Eq.~\eqref{Eq:AdiabaticCondition}]. }
\label{Fig:FewLevel}
\vspace{-10pt}
\end{figure} 

\begin{figure*}
\centering
\includegraphics[width=\textwidth]{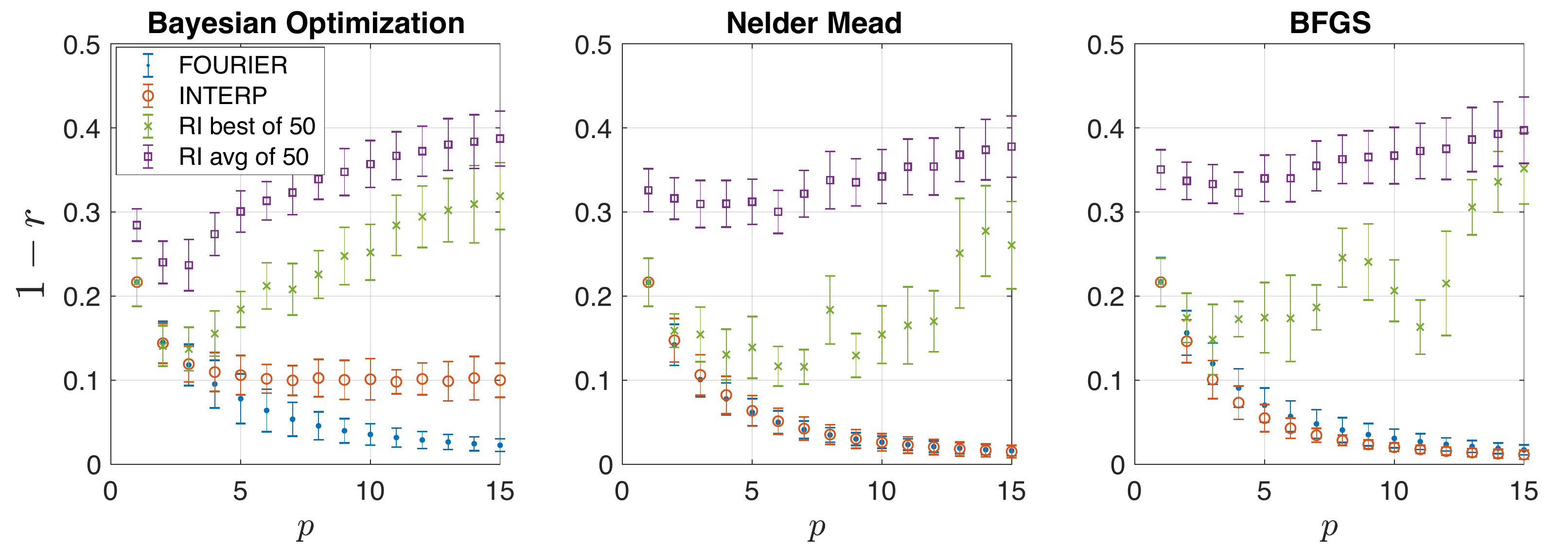}
\caption{\label{fig:diff-alg}
Comparison of three different optimization routines applied to 10 instances of 14-vertex w3R graphs. The initial point of optimization is generated with either our heuristics (FOURIER or INTERP) or random initialization (RI).
We plot the fractional error, $1-r$, averaged over instances, for each optimization routine and each initialization strategy at each $p$.
The error bars are sample standard deviations from the 10 instances.
For our heuristic strategy, the optimization starts with a single initial point generated using our FOURIER[$\infty,0$] or INTERP heuristics as described in Appendix~\ref{appx:heuristics}.
For the RI strategy, we generate 50 random initial points uniformly in the parameter space, and optimize from each initial point; both the best and the average of 50 RI runs are plotted.
}
\end{figure*}

In this Appendix, we elucidate the mechanism of the diabatic bump observed in Fig.~\ref{Fig:QAOAvsQA}(c) via an effective few-level dynamics. To study the intermediate dynamics during quantum annealing, we can work in the basis of instantaneous eigenstates $\ket{\epsilon_{l}(t)}$, where
\begin{equation}
H_{\text{QA}}(t) \ket{\epsilon_{l}(t)} = \epsilon_{l}(t) \ket{\epsilon_{l}(t)}.
\label{Eq:InstBasis}
\end{equation}
Expanding the time-evolved state in this basis as $\ket{\psi(t)} = \sum_{l} a_{l}(t) \ket{\epsilon_{l}(t)}$, the Schrodinger equation can be written as 
\begin{equation}
i \sum_{l} \left(\dot{a}_{l} \ket{\epsilon_{l}} + a_{l} \ket{\dot\epsilon_{l}} \right) = \sum_{l} \epsilon_{l} a_{l}  \ket{\epsilon_{l}},  
\end{equation}
where $\hbar = 1$ and the time dependence in the notations is dropped for convenience. Multiplying the equation by $\bra{\epsilon_{k}}$, the Schrodinger equation becomes 
\begin{equation}
i \dot{a}_{k} = \epsilon_{k} a_{k} - i\sum_{l \neq k} a_{l} \braket{\epsilon_{k}}{\dot{\epsilon}_{l}},  
\end{equation}
where $\braket{\epsilon_{k}}{\dot{\epsilon}_{k}} = 0$ is taken by absorbing the phase into the eigenvector $\ket{\epsilon_{k}}$. Written in a matrix form, we have 
\begin{align}
i \left(\begin{array}{c}\dot{a}_0 \\\dot{a}_1 \\\dot{a}_2 \vspace{-.1cm}\\ \vdots \end{array}\right) \!  
= \! 
\left( \! \! \begin{array}{cccc}
0 & -i \braket{\epsilon_0}{\dot{\epsilon}_1} & -i \braket{\epsilon_0}{\dot{\epsilon}_2} & \! \! \cdots  \\
-i \braket{\epsilon_1}{\dot{\epsilon}_0} & \Delta_{10} & -i \braket{\epsilon_1}{\dot{\epsilon}_2} & \! \! \cdots  \\
-i \braket{\epsilon_2}{\dot{\epsilon}_0} & -i \braket{\epsilon_2}{\dot{\epsilon}_1} & \Delta_{20} & \! \! \cdots   \vspace{-.1cm} \\
 \vdots & \vdots  & \vdots & \! \! \ddots   \end{array} \! \right) 
\! \!   \left(\begin{array}{c} a_0 \\ a_1 \\ a_2    \vspace{-.1cm} \\ \vdots \end{array}\right), 
\label{Eq:InstBasisDynamics}
\end{align}
where we take the ground state energy $\epsilon_{0} = 0$ (by absorbing it into the phase of the coefficients) and $\Delta_{i0} = \epsilon_{i} - \epsilon_{0}$ is the instantaneous energy gap from the $i$th excited state to the ground state. The time evolution starts from the initial ground state with $a_{0} = 1$ and $a_{i} = 0$ for $i \neq 0$, and the adiabatic condition to prevent coupling to excited states is 
\begin{equation} \label{Eq:AdiabaticCondition}
\dfrac{\left|\braket{\epsilon_0}{\dot{\epsilon}_i} \right|}{\Delta_{i0}} =  \dfrac{\left|\bracket{\epsilon_0}{\frac{dH_{\text{QA}}}{dt}}{\epsilon_{i}} \right|}{\Delta_{i0}^{2}} = \dfrac{\left|\bracket{\epsilon_0}{\frac{dH_{\text{QA}}}{ds}}{\epsilon_{i}} \right|}{\Delta_{i0}^{2} T} \ll 1. 
\end{equation}
The first equality can be derived from Eq.~\eqref{Eq:InstBasis}. This produces the standard adiabatic condition $T = O(1/\Delta_{\min}^{2})$. As we discussed in section \ref{subSec:mechanism}, the minimum gap for some graphs can be exceedingly small, so the adiabatic limit is not practical. However, it may be possible to choose an appropriate run time $T$, which breaks adiabaticity, but is long enough such that only few excited states are effectively involved in the dynamics. This is the regime where the diabatic bump operates and one can understand the dynamics by truncating Eq.~\eqref{Eq:InstBasisDynamics} to the first few basis states.

As an example, we plot in Fig.~\ref{Fig:FewLevel}(a) the instantaneous eigenstate populations of the first few states. It is simulated with the full Hilbert space, but effectively the same dynamics will be generated if the simulation is restricted to the first few basis states in Eq.~\eqref{Eq:InstBasisDynamics}. Fig.~\ref{Fig:FewLevel}(b) shows the strength of the couplings between the instantaneous ground state and the low excited states. By comparing Fig.~\ref{Fig:FewLevel}(a) and Fig.~\ref{Fig:FewLevel}(b), one can see that $ T = T^{*} = 40$ allows the time evolution to break the adiabatic condition before the anticrossing: population leaks to the first excited state, which becomes the ground state after the anticrossing. Thus, the time scale of $T^{*}$ for the diabatic bump represents a delicate balance between allowing population to leak out of the ground state and suppressing excessive population leakage, which explains why it happens at a certain range of time scale.

\section{Comparing different classical optimization routines \label{appx:diff-alg}}
In this appendix, we compare three different classical routines that can be used to optimize QAOA parameters: Bayesian Optimization~\cite{bayesopt}, Nelder-Mead~\cite{NelderMead}, and BFGS~\cite{BFGS1, *BFGS2, *BFGS3, *BFGS4}.
This comparison is done by a numerical experiment where we apply these optimization routines to 10 instances of 14-vertex weighted 3-regular (w3R) graphs.
To compare them on equal footing, we terminate each optimization run after a budget of $20p$ objective function evaluations is used.
In the gradient-based routine, BFGS, we include the cost of gradient estimation via the finite-difference method into the budget of $20p$ objective function evaluations.
For each routine, we start at $p=1$ and gradually increment $p$, and perform optimization where the initial point is generated using either our heuristic strategies (FOURIER and INTERP) or the standard strategy of random initialization (RI).
We use the versions of these optimization routines implemented in MATLAB R2017b as \texttt{bayesopt}, \texttt{fminsearch}, and \texttt{fminunc}, respectively.
The objective function at each set of parameter is evaluated to floating point precision.
The tolerance in both objective function value and step size in parameter space, as well as the finite-difference-gradient step size, are chosen to be 0.01.

The result of our numerical experiment is plotted in Fig.~\ref{fig:diff-alg}.
Similar to Fig.~\ref{fig:heuristic}(a), we see that the average quality of local optimum found from 50 RI runs is much worse than the best, indicating the difficulty of optimizing in the QAOA parameter landscape without a good initial point.
We also see, regardless of the classical routines chosen, one run of optimization from an initial point generated from our heuristic strategies is generally better than the best out of 50 runs from randomly generated initial points.
This indicates that our heuristic strategies work much better than RI and can be integrated with a variety of classical optimization routines.
Moreover, we find that Bayesian Optimization typically does not do as well as Nelder-Mead or BFGS for larger $p$, which is consistent with the folklore that this routine is better suited for low-dimensional parameter space~\cite{bayesopt}.
On the other hand, it seems both Nelder-Mead and BFGS have comparable performance, even when the cost of gradient evaluation is taken into account. The slight difference we observe between the two in Fig.~\ref{fig:diff-alg} is inconclusive and can be attributed to sub-optimal choices of tolerance and step size parameters and the deliberately limited budget of objective function evaluations.

\section{Details of Simulation with Measurement Projection Noise\label{appx:shot-noise}}

When running QAOA on actual quantum devices, the objective function is evaluated by averaging over many measurement outcomes, and consequently its precision is limited by the so-called measurement projection noise from quantum fluctuations.
We account for this effect by performing full Monte-Carlo simulations of actual measurements, where the simulated quantum processor only outputs approximate values of the objective function obtained by averaging $M$ measurements:
\begin{equation}
\bar{F}_{p,M} = \frac{1}{M} \sum_{i=1}^M C(\vect{z}_{p,i}),
\end{equation}
where $\z_{p,i}$ is a random variable corresponding to the $i$th measurement outcome obtained by measuring $\ket{\psi_p(\vgamma,\vbeta)}$ in the computational basis, and $C(\z)$ is the classical objective function.
Note when $M\to\infty$, we obtain $\bar{F}_{p,M}\to F_p=\bracket{\psi_p(\vgamma,\vbeta)}{H_C}{\psi_p(\vgamma,\vbeta)}$.
In the simulation, we achieve finite precision $|\bar{F}_{p,M}-F_p|\lesssim\xi$ by sampling measurements until the cumulative standard error of the mean falls below the target precision level $\xi$.
In other words, for each evaluation of $F_p$ requested by the classical optimizer, the number of measurements $M$ performed is set by the following criterion:
\begin{equation}
\sqrt{\frac{1}{M(M-1)}\sum_{i=1}^M[C(\vect{z}_{p,i})-\bar{F}_{p,M}]^2} \le \xi.
\end{equation}
Roughly, we expect $M\approx \Var(F_p)/\xi^2$.
To address issues with finite sample sizes, we also require that at least 10 measurements be performed ($M\ge 10$) for each objective function evaluation.

We now provide some details on the classical optimization algorithm used to optimize QAOA parameters.
Generally, classical optimization algorithms iteratively uses information from some given parameter point $(\vgamma,\vbeta)$ to find a new parameter point $(\vgamma',\vbeta')$ that hopefully produces a larger value of the objective function $F_p(\vgamma',\vbeta')\ge F_p(\vgamma,\vbeta)$.
In order for the algorithm to terminate, we need to set some stopping criteria.
Here, we specify two: First, we set an objective function tolerance $\epsilon$, such that if the change in objective function $|\bar{F}_{p,M'}(\vgamma',\vbeta') - \bar{F}_{p,M} (\vgamma,\vbeta)|\le \epsilon$, the algorithm terminates.
We also set a step-tolerance $\delta$, so that the algorithm terminates if the new parameter point is very close to the previous one $|\vgamma'-\vgamma|^2+|\vbeta'-\vbeta|^2\le \delta^2$.
For gradient-based optimization algorithms such as BFGS, we also use $\delta$ as the increment size for estimating gradients via the finite-difference method: $\partial F_p/\partial\gamma_i\simeq [\bar{F}_{p,M'}(\gamma_i+\delta)-\bar{F}_{p,M}(\gamma_i)]/\delta$.
In our simulations, we use the BFGS algorithm implemented  as \texttt{fminunc} in the standard library of MATLAB R2017b.

Using the approach described above, we simulate experiments of optimizing QAOA with measurement projection noise for a few example instances, with various choices of precision parameters $(\epsilon,\xi,\delta)$ and initial points.
For the representative instance studied in Fig.~\ref{Fig:ShotNoise}, we set $\epsilon = 0.1$, $\xi=0.05$, and $\delta=0.01$.
In each run of the simulated experiment, we start with QAOA of level either $p=1$ or $p=5$, and optimize increasing levels of QAOA using our FOURIER[$\infty,0$] heuristic strategy.
The initial point of QAOA optimization is either randomly selected (when starting at $p=1$) or chosen based on an educated guess using optimal parameters from small-sized instances (at $p=1$ and $p=5$).
Specifically, the educated guess for the initial points we use are $(\vu^0,\vv^0)=(1.4849,0.5409)$ at level $p=1$, and
\begin{align}
\vu^0 &= (1.9212,0.2891,0.1601,0.0564,0.0292), \\
\vv^0 &= (0.6055,-0.0178,0.0431,-0.0061,0.0141)
\end{align}
at level $p=5$.
For each such run, we keep track of the history of all the measurements, so that the largest cut $\text{Cut}_i$ found after the $i$-th measurement can be calculated.
We repeat each experiment for 500 times with different pseudo-random number generation seeds, and average over their histories.

\section{Techniques to speed up numerical simulation}

In this Appendix, we discuss a number of techniques we exploited to speed up the numerical simulation for both QAOA and QA. 

First, we make use of the symmetries present in the Hamiltonian. For general graphs, the only symmetry operator that commutes with both $H_{C}$ and $H_{B}$ is the parity operator $\mathcal{P} = \prod_{i=1}^{N} \sigma_{i}^{x}$: $[H_{C}, \mathcal P] = 0, [H_{B}, \mathcal P] = 0$, and so does  $[H_{\text{QA}}(s), \mathcal P] =0$, where $H_{\text{QA}}(s)$ is the quantum annealing Hamiltonian in Eq.~\eqref{Eq:HamQA}. The parity operator has two eigenvalues, $+1$ and $-1$, each with half of the entire Hilbert space. The initial state for both QAOA and QA is in the positive sector, i.e., $\mathcal P \ket{+}^{\otimes N} = \ket{+}^{\otimes N}$. Thus, any dynamics remain in the positive parity sector. We can rewrite $H_{C}$ and $H_{B}$ in the basis of the eigenvectors of $\mathcal P$, and reduce the Hilbert space from $2^{N}$ to $2^{N-1}$ by working in the positive parity sector. 

For QA, dynamics involving the time-dependent Hamiltonian can be simulated by dividing the total simulation time $T$ into sufficiently small discrete time $\tau$ and implement each time step sequentially. At each small step, one can evolve the state without forming the full evolution operator~\cite{Moler2003Nineteen}, either using Krylov subspace projection method \cite{Sidje1998Expokit} or a truncated Taylor series approximation \cite{AlMohy:2011iw}. In our simulation, we used a scaling and squaring method with a truncated Taylor series approximation \cite{AlMohy:2011iw} as it appears to run slightly faster than the Krylov subspace method for small time steps. 

For QAOA, the dynamics can be implemented in a more efficient way due to the special form of the operators $H_{C}$ and $H_{B}$. We work in the standard $\sigma_{z}$ basis. Thus, $H_{C} = \sum_{\left<i,j \right>} \frac{w_{i j}}{2} (\Id - \sigma_{i}^{z} \sigma_{j}^{z})$ can be written as a diagonal matrix and the action of $e^{-i \gamma H_{C}}$ can be implemented as vector operations. For $H_{B}$, the time evolution operator can be simplified as 
\begin{equation}
e^{-i \beta H_{B}} = \prod_{j=1}^{N} e^{-i \beta \sigma_{j}^{x}} = \prod_{j=1}^{N} \left( \Id\cos \beta  - i  \sigma_{j}^{x} \sin \beta \right).
\end{equation}
Therefore, the action of $e^{-i \beta H_{B}}$ can also be implemented as $N$ sequential vector operations without explicitly forming the sparse matrix $H_{B}$, which both improves simulation speed and saves memory. In addition, in the optimization of variational parameters, we calculated the gradient analytically, instead of using finite-difference methods. Techniques similar to the gradient ascent pulse engineering (GRAPE) method \cite{Khaneja2005Optimal} were used, which reduces the cost of computing the gradient from $O(p^{2})$ to $O(p)$, for a $p$-level QAOA. Lastly, in our FOURIER strategy, we need to calculate the gradient of the objective function with respect to the new parameters $(\vu,\vv)$. Since $\vgamma=A_S\vu$ and $\vbeta=A_C\vv$ for some matrices $A_S$ and $A_C$, their gradients are also related via $\nabla_{\vu} = A_S\nabla_{\vgamma}$ and $\nabla_{\vv} = A_S\nabla_{\vbeta}$.

\begin{figure*}
\includegraphics[width=\textwidth]{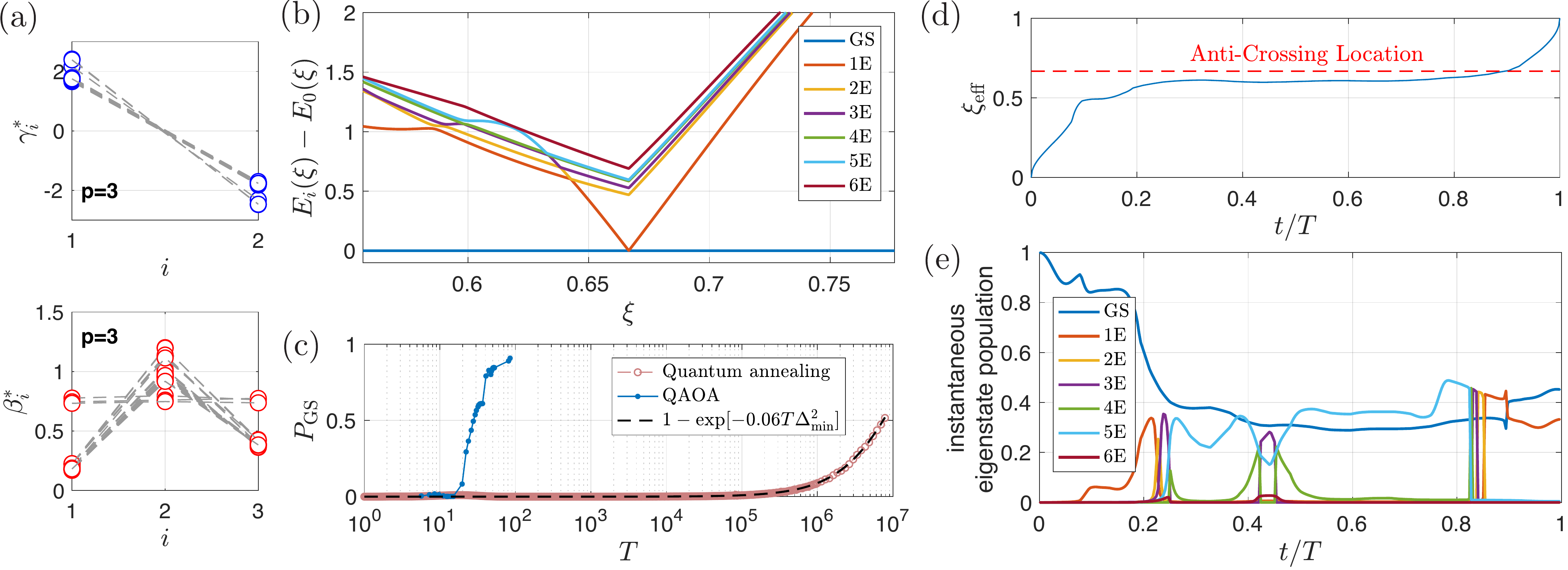}
\caption{\label{fig:MIS}(a) Pattern in the optimal QAOA parameters at level $p=3$ for 20 random instances of Maximum Independent Set (MIS). Each dashed line connects parameters for one particular graph instance.
(b) The energy difference relative to the ground state in the quantum annealing protocol of Eq.~\eqref{eq:QA-MIS} for an example 32-vertex MIS instance. The annealing protocol progresses as the parameter $\xi$ increases from 0 to 1. The minimum spectral gap between the ground state (GS) and the first excited state (1E) is $\Delta_{\min} = 0.0012$ at $\xi=0.6666$.
(c) Comparing performance of quantum annealing and QAOA on the example instance, in terms the ground state population at the end of the quantum evolution. The equivalent evolution time for QAOA is calculated via $T_\text{QAOA} = \sum_{i=1}^{p-1} |\gamma_i| + \sum_{i=1}^{p}|\beta_i|$.
(d) The effective annealing schedule converted from optimized 25-level QAOA parameters for the example MIS instance. 
(e) The population of the system in the instantaneous eigenstates, during the effective annealing schedule that approximates the dynamics under 25-level QAOA. Here, we observe that the algorithm attempts to transport the system to the fifth excited state, keeping it there before it undergoes a series of anti-crossings towards the ground state.
}
\end{figure*}

\vspace{-10pt}

\section{QAOA for Maximum Independent Set\label{sec:MIS}}

In this Appendix, we briefly illustrate the generality of our results by applying QAOA to another class of combinatorial optimization problems called Maximum Independent Set (MIS).
Given a graph $G=(V,E)$, the MIS problem concerns finding the largest independent set---a subset of vertices where no two of which share an edge.
In other words, the problem Hamiltonian is
\begin{equation}
H_P = \cP_{\rm IS}\left(\sum_i \hat{n}_i \right) \cP_{\rm IS}
\end{equation}
where $\hat{n} = \ketbra{1} = (\Id-\sigma^z)/2$, and $\cP_{\rm IS}$ is the projection (or restriction) onto the subspace of independent set states $\text{span}\{\ket{\psi}: \hat{n}_i\hat{n}_j \ket{\psi} = 0~ \forall (i,j)\in E\}$.
The $p$-level QAOA for MIS, first suggested by \cite{QAOA}, involves the preparation of the following variational wavefunction
\begin{equation}
\vspace{-5pt}
\ket{\psi_p(\vgamma,\vbeta)} = e^{-i\beta_p H_Q} \prod_{k=1}^{p-1} e^{-i\gamma_k H_P} e^{-i\beta_k H_Q}\ket{0}^{\otimes N},
\end{equation}
where
\begin{equation}
H_Q = \cP_{\rm IS} \left( \sum_i \sigma_i^x \right)\cP_{\rm IS}.
\end{equation}
Similar to the case of MaxCut, here QAOA works by repeatedly measuring the system in the computational basis to obtain an estimate of $G_p(\vgamma, \vbeta) = \bracket{\psi_p(\vgamma, \vbeta)}{H_P}{\psi_p(\vgamma, \vbeta)}$,
and using a classical computer to search for the best variational parameters $(\vgamma^*, \vbeta^*)$ so as to maximize $G_p$.
We note the evolution can be implemented using the following physical Hamiltonian
\begin{equation}
H^{\rm MIS}_{\rm physical}(t) = \sum_i [\Delta(t) \hat{n}_i+ \Omega(t)\sigma_i^x] + \sum_{\ave{i,j}} U \hat{n}_i \hat{n}_j.
\end{equation}
In the $U\gg |\Delta|, |\Omega|$ limit, the system is constraint to the manifold where $\hat{n}_i \hat{n}_j = 0$ for all edges $\ave{i,j}$ (since the initial state is $\ket{0}^{\otimes N}$), and the QAOA circuit can be performed by setting appropriate waveforms of $\Delta(t)$ and $\Omega(t)$. See Ref.~\cite{Pichler2018Quantum} for an implementation scheme of MIS with a platform of neutral atoms interacting via Rydberg excitations.

After performing exhaustive search of QAOA parameters using randomly initialized optimization for many instances of MIS, we discover a similar pattern.
This is illustrated in Fig.~\ref{fig:MIS}(a), where we see that the optimal parameters at $p=3$ tend to cluster in two visually distinct groups.
For one of the groups, the smooth curve underlying the parameters exhibit resemblance to a quantum annealing protocol, using a time-dependent Hamiltonian:
\begin{equation}
H_{\rm QA}^{\rm MIS}(t) = f_P(t) H_P + f_Q(t) H_Q.
\end{equation}
For example, if we choose $(f_P, f_Q)$ such that $f_P(0) > 0$ and $f_P(T) < 0$, and $f_Q(0)=f_Q(T)=0$,
 then we can initialize the system in $\ket{0}^{\otimes N}$, which is the ground state of  $H_{\rm QA}^{\rm MIS}(t=0)$, and evolve adiabatically to reach the ground state of $H_{\rm QA}^{\rm MIS}(t=T)\propto -H_P$ (i.e. state encoding the MIS solution).
 The Hamiltonian $H_{Q}$, which is turned on in the middle of the time evolution, induces couplings between different independent-set basis states and opens a spectral gap between the ground state and excited states.  
For concreteness of discussion, we focus on the following choice of annealing schedule:
\begin{equation} \label{eq:QA-MIS}
f_P^c(\xi) = 6 (1-2\xi), ~ f_Q^c(\xi) = \sin^2 (\pi \xi), ~ \xi(t) = t/T.
\end{equation}


We further analyze the performance and mechanism of QAOA for MIS by focusing on example instances that are difficult for adiabatic quantum annealing due to small spectral gaps.
In Fig.~\ref{fig:MIS}(b), we show the level-crossing structure for such an example instance, where the minimum spectral gap is $\Delta_{\min} = 0.0012$. 
The same instance is studied in Ref.~\cite{Pichler2018Quantum}. 
To study the performance of QAOA in deeper-depth circuits, we use the interpolation-based heuristic strategy outlined in Appendix~\ref{appx:INTERP} to optimize QAOA parameters for this example instance starting at level $p=3$.
The performance of QAOA and quantum annealing are then compared in Fig.~\ref{fig:MIS}(c), where we see that QAOA is able to obtain a much larger ground state population in much shorter time compared to the adiabatic time scale of $1/\Delta_{\min}^2 \approx 10^6$.
We then study the mechanism of QAOA by converting its parameters at level $p=25$ to a smooth annealing path $(f_P^{\rm QAOA}, f_Q^{\rm QAOA})$ in a similar procedure as in Sec.~\ref{subSec:mechanism}.
This annealing path is visualized in Fig.~\ref{fig:MIS}(d), where we plot the effective $\xi_\text{eff}(t)$ defined by $f_P^c(\xi_\text{eff}(t))/f_Q^c(\xi_\text{eff}(t)) = f_P^{\rm QAOA}(t)/f_Q^{\rm QAOA}(t)$ for the QAOA-like schedule.
We then monitor populations in the instantaneous eigenstates during the evolution to gain insights into the mechanism of QAOA.
As shown in Fig.~\ref{fig:MIS}(e), QAOA is able to learn to navigate a very complicated level-crossing structure by a combination of adiabatic and non-adiabatic operations: the system diabatically couples to the excited states, lingers to maximize population in the fifth excited state, and then exploits a series of anti-crossings to return to the ground state.
Our results here demonstrate that the non-adiabatic mechanisms observed in QAOA for MaxCut can play a significant role in more general problems, such as difficult MIS instances.

\end{document}